\RequirePackage{fix-cm}
\RequirePackage{amsmath}

\documentclass[smallextended]{svjour3}       % onecolumn (second format)
\smartqed  % flush right qed marks, e.g. at end of proof
\usepackage{graphicx}
\usepackage{amsmath}
\usepackage{hyperref}
\usepackage{amssymb}

\usepackage{mathptmx}  
\journalname{Journal of Statistical Physics}
\begin{document}

\title{Dynamical large deviations for homogeneous systems with long range
interactions and the Balescu--Guernsey--Lenard equation%\thanks{Grants or other notes
}

\titlerunning{Dynamical large deviations for particles with long range
interactions}        % if too long for running head

\author{Ouassim Feliachi 
 and Freddy Bouchet       %etc.
}


\institute{
              O. Feliachi \at
Institut Denis Poisson, Université d’Orléans, CNRS, Université de Tours, France\\             
Univ Lyon, Ens de Lyon, Univ Claude Bernard, CNRS, Laboratoire de Physique, Lyon, France \\     
            \email{ouassim.feliachi@univ-orleans.fr}             %  \\
               \and         
           F. Bouchet \at
              Univ Lyon, Ens de Lyon, Univ Claude Bernard, CNRS, Laboratoire de Physique, Lyon, France \\
              \email{freddy.bouchet@ens-lyon.fr}           %  \\
}

\date{Received: date / Accepted: date}

\maketitle

\begin{abstract}
We establish a large deviation principle for time dependent trajectories (paths) of the empirical density
of $N$ particles with long range interactions, for homogeneous systems.
This result extends the classical kinetic theory that leads to the
Balescu--Guernsey--Lenard kinetic equation, by the explicit computation
of the probability of typical and large fluctuations. The large deviation
principle for the paths of the empirical density is obtained through
explicit computations of a large deviation Hamiltonian. This Hamiltonian
encodes all the cumulants for the fluctuations of the empirical density,
after time averaging of the fast fluctuations. It satisfies a time
reversal symmetry, related to the detailed balance for the stochastic
process of the empirical density. This explains in a very simple way
the increase of the macrostate entropy for the most probable states,
while the stochastic process is time reversible, and describes the
complete stochastic process at the level of large deviations. 
\end{abstract}

\section{Introduction}

We consider the Hamiltonian dynamics of particles that interact through a mean-field potential. The dynamics
reads 
\begin{equation}
\begin{cases}
{\displaystyle \frac{\text{d}\mathbf{r}_{n}}{\text{d}t}} & =\mathbf{v}_{n}\\
\\
{\displaystyle \frac{\text{d}\mathbf{v}_{n}}{\text{d}t}} & {\displaystyle =-\frac{1}{N}\sum_{m\neq n}\frac{\text{d}}{\text{d}\mathbf{r}_{n}}W\left(\mathbf{r}_{n}-\mathbf{r}_{m}\right)}
\end{cases}\label{eq:dynamics-1}
\end{equation}
where $\left\{ \mathbf{r}_{n}\right\} _{1\leq n\leq N}$ are the positions
and $\left\{ \mathbf{v}_{n}\right\} _{1\leq n\leq N}$ the velocities.
This set-up is relevant for plasmas in the weak coupling regime \cite{Nicholson_1991},
self-gravitating systems \cite{Padmanabhan_1990,Chavanis_2006IJMPB_Revue_Auto_Gravitant},
and many particle systems with long range interactions \cite{Bouchet_Gupta_Mukamel_PRL_2009}.
It also shares many theoretical analogies with two-dimensional and
geostrophic turbulence, through the point vortex model \cite{Chavanis_2001PhRvE_64_PointsVortex,Kiessling_Lebowitz_1997_PointVortex_Inequivalence_LMathPhys},
or stochastically forced models of two-dimensional and geostrophic
turbulence \cite{Bouchet_Nardini_Tangarife_2013_Kinetic_JStatPhys}.
The kinetic theory of systems with mean-field potentials (or long
range interactions) is a classical piece of theoretical physics. The
relaxation to equilibrium of the empirical density 
\[
g_{N}(\mathbf{r},\mathbf{v},t)=\frac{1}{N}\sum_{n=1}^{N}\delta(\mathbf{r}-\mathbf{r}_{n}(t))\delta(\mathbf{v}-\mathbf{v}_{n}(t)),
\]
is described by the Balescu--Guernsey--Lenard kinetic equation in
the limit of a large number of particles. This result has been formally
derived by Balescu, Guernsey and Lenard \cite{balescu1960irreversible,balescu1963kinetic}.
In the context of plasma physics where we consider $N$ charged particles
submitted to Coulomb interactions, we refer to Nicholson \cite{Nicholson_1991}
for a derivation using the BBGKY hierarchy, or to Lifshitz and Pitaevskii
\cite{Lifshitz_Pitaevskii_1981_Physical_Kinetics} who follow the
Klimontovich approach.

In this paper we extend this classical kinetic theory by describing
the statistics of the large deviations for time dependent trajectories (paths) of the
empirical density. For simplicity, we restrict our analysis to paths
of the empirical density which remain close to homogenous distributions.
We consider the projection of the empirical density on homogeneous
distributions: $f_{N}\left(\mathbf{v},t\right)=N^{-1}L^{-3}\int\text{d}\mathbf{r}\,g_{N}\left(\mathbf{r},\mathbf{v},t\right)=N^{-1}L^{-3}\sum_{n=1}^{N}\delta(\mathbf{v}-\mathbf{v}_{n}(t))$,
where $L^{3}$ is the volume of the system. The natural evolution
of $f_{N}$ occurs on time scales of order $N$ (except in dimension
$d=1$ \cite{Yamaguchi_Barre_Bouchet_DR:2004_PhysicaA}). After time
rescaling $\tau=t/N$, we study the probability of $f_{N}^{s}\left(\mathbf{v},\tau\right)=f_{N}\left(\mathbf{v},N\tau\right)$
(by abuse of notation and for convenience, we still denote $f_{N}^{s}=f_{N}$).
We justify that the probability that a path $\left\{ f_{N}(\tau)\right\} _{0\leq\tau\leq T}$
remains in the neighborhood of a prescribed path $\left\{ f(\tau)\right\} _{0\leq t\leq T}$
satisfies the large deviation principle
\begin{equation}
\mathbf{P}\left(\left\{ f_{N}(\tau)\right\} _{0\leq\tau\leq T}=\left\{ f(\tau)\right\} _{0\leq\tau\leq T}\right)\underset{N\rightarrow\infty}{\asymp}\text{e}^{-NL^{3}\int_{0}^{T}\text{d}\tau\,\text{Sup}_{p}\left\{ \int\text{d}\mathbf{v}\,\dot{f}p-H[f,p]\right\} }\text{e}^{-NI_{0}\left[f^0\right]},\label{eq:Large_Deviation_Principle_Intro}
\end{equation}
where $\dot{f}$ is the time derivative of $f$, $p$ is a function
over the velocity space and is called the conjugated momentum of $\dot{f}$,
the Hamiltonian $H$ is a functional of $f$ and $p$ that characterizes
the dynamical fluctuations, $I_{0}$ is a large deviation rate
function for the initial conditions of $f_{N}$, and where the symbol
$\underset{N\rightarrow\infty}{\asymp}$ means a large deviation principle
(roughly speaking, the log of the left hand side is equivalent to
the log of the right hand side, see classical textbooks \cite{Dembo_Zeitouni_1993_Book}
for a more precise definition). We note that $H$ is not the Hamiltonian
of the microscopic dynamics but it rather defines a statistical
field theory that quantifies the probabilities of paths of the empirical
density. 

The main result of this paper is the first computation of an explicit
expression for $H$ and the study of its symmetry properties. The
explicit expression for $H$ is 
\begin{equation}
H\left[f,p\right]=-\frac{1}{4\pi L^{3}}\sum_{\mathbf{k}}\int\text{d}\omega\,\log\left\{ 1-\mathcal{J}\left[f,p\right]\left(\mathbf{k},\omega\right)\right\} ,\label{eq:LDP_Hamiltonien_Intro}
\end{equation}
with
\begin{multline}
\mathcal{J}\left[f,p\right]\left(\mathbf{k},\omega\right)=4\pi\int\text{d}\mathbf{v}_{1}\text{d}\mathbf{v}_{2}\,\frac{\partial p}{\partial\mathbf{v}_{1}}\cdot\mathbf{A}\left[f\right]\left(\mathbf{k},\omega,\mathbf{v}_{1},\mathbf{v}_{2}\right)\cdot\left\{ \frac{\partial f}{\partial\mathbf{v}_{2}}f(\mathbf{v}_{1})-f(\mathbf{v}_{2})\frac{\partial f}{\partial\mathbf{v}_{1}}\right\} \\
+4\pi\int\text{d}\mathbf{v}_{1}\text{d}\mathbf{v}_{2}\,\left\{ \frac{\partial p}{\partial\mathbf{v}_{1}}\frac{\partial p}{\partial\mathbf{v}_{1}}-\frac{\partial p}{\partial\mathbf{v}_{1}}\frac{\partial p}{\partial\mathbf{v}_{2}}\right\} :\mathbf{A}\left[f\right]\left(\mathbf{k},\omega,\mathbf{v}_{1},\mathbf{v}_{2}\right)f(\mathbf{v}_{1})f(\mathbf{v}_{2}),\label{eq:LDP_J_Intro}
\end{multline}
where the $\mathbf{A}\left(\mathbf{k},\omega,\mathbf{v}_{1},\mathbf{v}_{2}\right)$
is a symmetric tensor which can be expressed through the Fourier transform
of the interaction potential and the dielectric function.

Equations (\ref{eq:LDP_Hamiltonien_Intro}-\ref{eq:LDP_J_Intro})
clearly show that the Hamiltonian is not quadratic in the conjugated
momentum $p$. This shows that the fluctuations that lead to large
deviations are not locally Gaussian, by contrast with many other cases,
for instance when diffusive limits are involved as in the case of
macroscopic fluctuation theory \cite{bertini2015macroscopic}, or for
plasma fluctuations at scales much smaller than the Debye length \cite{2021arXiv210104455F}.
It is striking that it is possible to get explicit formulas (\ref{eq:LDP_Hamiltonien_Intro}-\ref{eq:LDP_J_Intro})
for the large deviation Hamiltonian, for which cumulants of all order
are relevant and non trivial. The four key theoretical ideas and technical
tools we use are: making the connection with large deviation theory
for slow-fast systems and identifying the statistics of the fast motion,
expressing the Hamiltonian as a functional determinant on a space
of functions that depend both on time and velocity, using the Szegö--Widom
theorem to reduce this functional determinant to a simpler one on
a space of functions that depend on velocity only, and finally computing
explicitly those determinants on the space of functions that depend
on velocity only. 

This article is the last of a series of three aimed at establishing
dynamical large deviation principles related to the main classical
kinetic theories, starting from Hamiltonian dynamics. The first one
\cite{bouchet2020boltzmann} dealt with the large deviations for dilute
gases (associated with the Boltzmann equation), the second one dealt
with large deviations for plasma fluctuations at scales much smaller
then the Debye length \cite{2021arXiv210104455F} (associated with
the Landau equation), and this one deals with the case of systems
with mean-field interactions (associated with the Balescu--Guernsey--Lenard
equation). The key point of these three works is to establish large
deviation principles for particle systems with Hamiltonian dynamics. At first sight it might seem surprising to obtain a stochastic process for an effective kinetic description, starting from a deterministic dynamics. However it is well known that, after taking the limit with an infinite time scale separation between the slow and fast degrees of freedom, the effective dynamics of a slow-fast dynamical system, with chaotic fast degrees of freedom, is stochastic. At the level of large deviations, for deterministic dynamical systems, mathematicians have proven theorems that establish large deviation principles for the effective stochastic process of the slow variable, from natural hypotheses~\cite{kifer1992averaging,kifer2004averaging}. This behavior can also be illustrated numerically, for instance coupling a slow dynamics with a fast chaotic Lorenz model dynamics~\cite{Melbourne_2011}.   
As far as we know, our three works (\cite{bouchet2020boltzmann,2021arXiv210104455F} and this one) and \cite{bodineau2020fluctuation}
establish the first large deviation principles, in kinetic theory
that do not start from stochastic dynamics, like for instance in macroscopic
fluctuation theory \cite{bertini2015macroscopic}. While in \cite{bodineau2020fluctuation} the result is proven for dilute gases in the Boltzmann--Grad limit for times of order of one collision times, our derivations
are not mathematical proofs. All the steps of our derivation
are however exact computations, once natural hypothesis are made,
in the spirit of the most precise classical works by theoretical physicists
in kinetic theory, and the result is expected to be valid for times much large that the kinetic times.

The large deviation results of the present paper (\ref{eq:LDP_Hamiltonien_Intro}-\ref{eq:LDP_J_Intro})
have some direct relations with the large deviations for plasma fluctuations
at a scale much smaller than the Debye length, considered in \cite{2021arXiv210104455F}.
While the result in the present paper is much more general, we explain
in section \ref{subsec:Gaussian-Approximation} that for plasma fluctuations
for scales much smaller than the Debye length, the locally Gaussian large deviation
Hamiltonian of \cite{2021arXiv210104455F} can be recovered. We also
prove in this paper a conjecture made in \cite{2021arXiv210104455F}
about the structure of the cumulant expansion. We stress that the
results of \cite{2021arXiv210104455F} are not just a sub-case of
the results in the present paper. There, we were also considering
questions of a different nature: the relation with the large deviation
Hamiltonian for dilute gases, on one hand, and the relation with large
deviations for effective mean-field diffusions, on the other hand.

We stress that our large deviation principle for paths immediately
implies a gradient flow structure for the Balescu--Guernsey--Lenard
operator, adapting to this specific case the general connection between path large deviation and gradient flows first
discussed in \cite{mielke2014relation} and simply explained in section
5 of \cite{bouchet2020boltzmann}. As far as we know, no gradient
flow structure was known before for the Balescu--Guernsey--Lenard
operator. As the large deviations are not locally Gaussian, this gradient structure is not a standard one, but a generalized one~\cite{mielke2014relation} (please see section
5 of \cite{bouchet2020boltzmann} for a simple definition standard of non standard gradient flow structure). \\

In parallel to our results, many mathematical results have recently
been obtained for the kinetic theory of plasma and systems with long
range interactions. The derivation of the Vlasov equation from the
$N$ particle dynamics had been first proved by Neunzert \cite{neunzert1977vlasov}, Braun and Hepp \cite{Braun_Hepp_CommMathPhys_1977} and Dobrushin \cite{dobrushin1979vlasov}, for interactions through
a smooth potential. This question is still under study for interaction potentials
with singularities, for instance with the Coulomb interaction (see
for instance \cite{Hauray_Jabin_ARMA_2007}). Kiessling's review \cite{kiessling2008microscopic}
provides a recent report on the mathematical justification of the
Vlasov equation from the microscopic dynamics of interacting particles.
The stability of stationary states of the Vlasov equation, for describing
the dynamics of the empirical density over time scales that diverge
with $N$, but which are much smaller than the kinetic time, has been
proven in \cite{Caglioti_Rousset_2007_JStatPhys_QSS}. The description
of Gaussian fluctuations of the potential, for dynamics close to the
Vlasov equilibrium, has been established by Braun and Hepp for smooth
interaction potentials, or in the book~\cite{Spohn_1991}.
More recent works \cite{lancellotti2009fluctuations,lancellotti2010vlasov,lancellotti2016time,paul2019size,velazquez2018two,duerinckx2021size}
discuss the Gaussian process of the fluctuations of the potential
close to a Vlasov solution. A recent proof has been proposed
for the the validity of the Balescu--Guernsey--Lenard equation up
to time scales of order $N^{r}$ with $r<1$ \cite{duerinckx2021lenard}. 

We also stress that the mathematics community is also very active
in studying large deviation principles related to kinetic theories.
Rezakhanlou pioneered large deviation principle results, related to
the Boltzmann equations \cite{rezakhanlou1998large}, in the case
of stochastic toy models of collisions. The first mathematical result
which deals with Hamiltonian dynamics, more specifically hard sphere
dynamics, has been obtained by Bodineau, Gallagher, Saint-Raymond,
and Simonella \cite{bodineau2020fluctuation,bodineau2020statistical}.
They basically prove the validity of the large deviation Hamiltonian
for times of order one collision time, in the spirit of Lanford's proof
for the Boltzmann equation. Large deviation principles have also been
obtained for Kac's models \cite{heydecker2021large,basile2021large}.
As far as we know, no mathematical results exists for the large deviations
of plasma or systems with mean-field interactions.\\

In section \ref{sec:Dynamics-of-plasma} we define the Hamiltonian
dynamics, as well as the classical kinetic equation that describes
the relaxation to equilibrium. In section \ref{sec:Computation-of-the},
we establish a large deviation principle for the empirical density
using the slow-fast decomposition of the quasilinear dynamics. In
section \ref{sec:Computation-of-the}, we provide an explicit computation
of the large deviation Hamiltonian. In section \ref{sec:Properties-of-the},
we check that this Hamiltonian is fully compatible with the conservation
laws of the system, as well as its time-reversal symmetry, and that it is 
consistent with statistics in the microcanonical ensemble. We discuss
perspectives in section \ref{sec:Perspectives}.

\section{Dynamics of particles with long range interactions\label{sec:Dynamics-of-plasma}}

In this section we set up the definitions, and present classical results
about the kinetic theory of the dynamics of $N$ particles with long
range interactions, in the limit of large $N$. In section \ref{subsec:Hamiltonian-dynamics},
we define the Hamiltonian dynamics. In section \ref{subsec:Vlasov-equation},
we introduce the Vlasov equation that describes the evolution of the
empirical density on timescales of order one. In section \ref{subsec:The-Balescu=002013Guernsey=002013Lenard-equa},
we introduce the Balescu--Guernsey--Lenard equation that describes
the long time relaxation of the empirical density, from Vlasov stationary
solutions to the Maxwell-Boltzmann equilibrium distribution, and some
of its important physical properties.

\subsection{Hamiltonian dynamics of $N$ particles with long range interactions\label{subsec:Hamiltonian-dynamics}}

We consider $N$ particles with positions $\left\{ \mathbf{r}_{n}\right\} _{1\leq n\leq N}$
and velocities $\left\{ \mathbf{v}_{n}\right\} _{1\leq n\leq N}$
governed by a Hamiltonian dynamics
\begin{equation}
\begin{cases}
{\displaystyle \frac{\text{d}\mathbf{r}_{n}}{\text{d}t}} & =\mathbf{v}_{n}\\
\\
{\displaystyle \frac{\text{d}\mathbf{v}_{n}}{\text{d}t}} & {\displaystyle =-\frac{1}{N}\sum_{m\neq n}\frac{\text{d}}{\text{d}\mathbf{r}_{n}}W\left(\mathbf{r}_{n}-\mathbf{r}_{m}\right)}
\end{cases}\label{eq:dynamics}
\end{equation}
where the interaction potential $W(\mathbf{r})$ is an even function
of $\mathbf{r}$. In the following, we consider that $\mathbf{r}_{n}$
belongs to a 3-dimensional torus of size $L^{3}$, and $\mathbf{v}_{n}\in\mathbb{R}^{3}$. We stress that our results are actually valid for any space dimension $d>1$.
We assume that the potential $W$ is a long range potential: the decay
of $W$ is slow enough, so that the interaction is dominated by the
collective effects of the $N$ particle rather than by local effects.
In an infinite space this condition would be met if the potential
would be non-integrable, for instance in dimension $d$ it would decay
asymptotically like a power law $1/\left|\mathbf{r}\right|^{d}$ or
more slowly. This condition is met in many physical systems, for instance
self-gravitating systems or weak interacting plasma (with a large plasma parameter). For any finite $L$, the condition that the potential decays more slowly than $1/r^{d}$ is a sufficient condition for the potential to be long range.

We call $\mu-$space the $\left(\mathbf{r},\mathbf{v}\right)$ space.
The $\mu-$space is of dimension $6$. Let us define $g_{N}$ the
$\mu-$space empirical density for the positions and velocities of
the $N$ particles

\[
g_{N}(\mathbf{r},\mathbf{v},t)=\frac{1}{N}\sum_{n=1}^{N}\delta(\mathbf{r}-\mathbf{r}_{n}(t))\delta(\mathbf{v}-\mathbf{v}_{n}(t)).
\]
In the following, we will study the stochastic process of the asymptotic dynamics of $g_{N}$, 
as the number of particles $N$ goes to infinity.

\subsection{The Vlasov equation\label{subsec:Vlasov-equation}}

From equation (\ref{eq:dynamics}), one immediately obtains the Klimontovich
equation
\begin{equation}
\frac{\partial g_{N}}{\partial t}+\mathbf{v}\cdot\frac{\partial g_{N}}{\partial\mathbf{r}}-\frac{\partial V\left[g_{N}\right]}{\partial\mathbf{r}}\cdot\frac{\partial g_{N}}{\partial\mathbf{v}}=0,\label{eq:Klimontovich_eq}
\end{equation}
where $V[g_{N}](\mathbf{r},t)=\int\text{d}\mathbf{v}'\text{d}\mathbf{r}'W(\mathbf{r}-\mathbf{r}')g_{N}(\mathbf{r}',\mathbf{v}',t)$.
This is an exact equation for the evolution of $g_{N}$, if $W$ is
regular enough. For the Coulomb interaction, the formal equation (\ref{eq:Klimontovich_eq})
has to be interpreted carefully. In the following, we do not discuss
the divergences that might occur related to small scale interactions.
At a mathematic level, this would be equivalent to considering a potential
which is regularized at small scales, and smooth. The Klimontovich
equation (\ref{eq:Klimontovich_eq}) contains all the information
about the trajectories of the $N$ particles. We would like to build
a kinetic theory, that describes the stochastic process for $g_{N}$
at a mesoscopic level.

An important first result is that the sequence $\{g_{N}\text{\}}$
obeys a law of large numbers when $N\rightarrow+\infty$. More precisely,
if we assume that there is a set of initial conditions $\{g_{N}^{0}\text{\}}$
such that $\lim_{N\rightarrow+\infty}g_{N}^{0}\left(\mathbf{r},\mathbf{v}\right)=g^{0}\left(\mathbf{r},\mathbf{v}\right)$,
then over a finite time interval $t\in\left[0,T\right]$, $\lim_{N\rightarrow+\infty}g_{N}\left(\mathbf{r},\mathbf{v},t\right)=g\left(\mathbf{r},\mathbf{v},t\right)$
where $g$ solves the Vlasov equation
\begin{equation}
\frac{\partial g}{\partial t}+\mathbf{v}\cdot\frac{\partial g}{\partial\mathbf{r}}-\frac{\partial V\left[g\right]}{\partial\mathbf{r}}\cdot\frac{\partial g}{\partial\mathbf{v}}=0\,\,\,\text{with}\,\,\,g\left(\mathbf{r},\mathbf{v},t=0\right)=g^{0}\left(\mathbf{r},\mathbf{v}\right).\label{eq:vlasov}
\end{equation}
While the solution of the Klimontonvich equation is a distribution that carries the whole information about the positions and velocities of all the particles, the Vlasov equation describes the evolution of a continuous mesoscopic density for the same evolution.  As the Klimontovich and the Vlasov equations are formally the same,
this law of large numbers is actually a stability result for the Vlasov equation in a space
of distributions. Such a result has first been proven for smooth potentials
by Braun and Hepp \cite{Braun_Hepp_CommMathPhys_1977} and Neunzert
\cite{neunzert1977vlasov}. 

The Vlasov equation has infinitely many Casimir conserved quantities. As a consequence, it has an infinite
number of stable stationary states \cite{Yamaguchi_Barre_Bouchet_DR:2004_PhysicaA}. 
Any homogeneous distribution $g\left(\text{\ensuremath{\mathbf{r}}},\mathbf{v}\right)=f(\mathbf{v})$
is a stationary solution of the Vlasov equation. In the following,
we will consider dynamics close to any homogeneous $f$ which is a linearly stable stationary solution
of the Vlasov equation. This linear stability can be assessed by studying the dielectric susceptibility
$\varepsilon[f](\mathbf{k},\omega)$ \cite{Nicholson_1991,Lifshitz_Pitaevskii_1981_Physical_Kinetics},
defined by
\begin{equation}
\varepsilon[f]\left(\mathbf{k},\omega\right)=1-\hat{W}\left(\mathbf{k}\right)\int\text{d}\mathbf{v}\frac{\mathbf{k}.\frac{\partial f}{\partial\mathbf{v}}}{\mathbf{k}.\mathbf{v}-\omega-i\eta},\label{eq:dielectric_function}
\end{equation}
where $\hat{W}\left(\mathbf{k}\right)$ is the $\mathbf{k}$-th Fourier
component of the interaction potential: $\hat{W}\left(\mathbf{k}\right)=\int\text{d}\mathbf{r}\,\exp\left(-i\mathbf{k}.\mathbf{r}\right)W\left(\mathbf{r}\right)$.
Equation (\ref{eq:dielectric_function}) and every other equations
involving $\pm i\eta$ have to be understood as the limit as $\eta$
goes to zero with $\eta$ positive. The dielectric susceptibility
function $\varepsilon$ plays the role of a dispersion relation in
the linearized dynamics, and a solution $f$ is stable if $\varepsilon[f]$
has no zeros except for $\omega$ on the real line. We note that $\varepsilon\left[f\right]\left(-\mathbf{k},-\omega\right)=\varepsilon^{*}\left[f\right]\left(\mathbf{k},\omega\right)$.
Another important property of the dielectric susceptibility is $\varepsilon\left[I\left[f\right]\right]\left(\mathbf{k},-\omega\right)=\varepsilon^{*}\left[f\right]\left(\mathbf{k},\omega\right)$,
where $I\left[f\left(\mathbf{v}\right)\right]=f\left(-\mathbf{v}\right)$.
This last property, associated to the time-reversal symmetry of the
Hamiltonian dynamics, will be used in section \ref{subsec:Time-reversal-symmetry,-quasi-po}. In this section we have discussed the linear stability of stationary solutions of the Vlasov equation while \cite{Yamaguchi_Barre_Bouchet_DR:2004_PhysicaA} defines different other notions of stability.

From the point of view of dynamical systems, those homogeneous solutions
might be attractors of the Vlasov equation, with some sort of asymptotic
stability. At a linear level, this convergence for some of the observables,
for instance the potential, is called Landau damping \cite{Nicholson_1991,Lifshitz_Pitaevskii_1981_Physical_Kinetics}.
Such a stability might also be true for the full dynamics. Indeed
some non-linear Landau damping results have recently been proven \cite{Mouhot_Villani:2009}.

In the following we will study the dynamics of $g_{N}$, when its
initial condition is close to a homogeneous stable state $f(\mathbf{v})$.
On time scales of order one, the distribution is stable and remains
close to $f$ according to the Vlasov equation. However a slow evolution
occurs on a timescale $\tau$ of order $N$, in spaces of dimension
$d>1$. For this reason, such $f$ are called quasi-stationary states
\cite{Yamaguchi_Barre_Bouchet_DR:2004_PhysicaA}. In the following
section, we explain that this slow evolution is described by the Balescu--Guernsey--Lenard
equation for most initial conditions, or as a law of large numbers. 

As a conclusion, the Balescu--Guernsey--Lenard equation appears as a mesoscopic description of the solution of the Klimontovich equation, for homogeneous solutions, which is valid up to time scales of order $N$, while the Vlasov equation is valid only up to time scales of order one. The Balescu--Guernsey--Lenard equation is a crucial correction to the Vlasov equation close to homogeneous solutions. Indeed homogeneous solution have no evolution theourgh the Vlasov equation as they are stationary, while they have an evolution of order one over times scales of order $N$ through the Balescu--Guernsey--Lenard equation.  

\subsection{The Balescu--Guernsey--Lenard equation\label{subsec:The-Balescu=002013Guernsey=002013Lenard-equa}}

With the rescaling of time $\tau=t/N$, we expect a law of large numbers
in the sense that ``for almost all initial conditions'' the empirical
density $g_{N}$ converges to $f$, with $f$ that evolves according
to the Balescu--Guernsey--Lenard equation

\begin{equation}
\frac{\partial f}{\partial\tau}=\frac{\partial}{\partial\mathbf{v}}.\int\text{d}\mathbf{v}_{2}\,\mathbf{B}\left[f\right](\mathbf{v},\mathbf{v}_{2})\left(-\frac{\partial f}{\partial\mathbf{v}_{2}}f(\mathbf{v})+f(\mathbf{v}_{2})\frac{\partial f}{\partial\mathbf{v}}\right),\label{eq:BLG_eq}
\end{equation}
with 
\begin{equation}
\mathbf{B}\left[f\right](\mathbf{v}_{1},\mathbf{v}_{2})=\frac{\pi}{L^{3}}\int_{-\infty}^{+\infty}\text{d}\omega\,\sum_{\mathbf{k}\in\left(2\pi/L\right)\mathbb{Z}^{3}}\frac{\hat{W}\left(\mathbf{k}\right)^{2}\mathbf{k}\mathbf{k}}{\left|\varepsilon[f]\left(\mathbf{k},\omega\right)\right|^{2}}\delta\left(\omega-\mathbf{k}.\mathbf{v}_{1}\right)\delta\left(\omega-\mathbf{k}.\mathbf{v}_{2}\right),\label{eq:B_GLB}
\end{equation}
where \textbf{$\mathbf{k}\mathbf{k}$ }denotes the tensor product
$\mathbf{k}\otimes\mathbf{k}$. The tensor $\mathbf{B}$ is called
the collision kernel of the Balescu--Guernsey--Lenard equation (by
analogy with the Boltzmann equation).

A recent proof has been proposed for the the validity of the Balescu--Guernsey--Lenard
equation up to time scales $t$ of order $N^{r}$ with $r<1$ \cite{duerinckx2021lenard}.
We know no mathematical proof of such a result for time scales $t$
of order $N$ ($\tau$ of order one). In the theoretical physics literature,
this equation is derived as an exact consequence of the dynamics once
natural hypotheses are made. Two classes of derivations are known,
either the BBGKY hierarchy detailed in \cite{Nicholson_1991} or the
Klimontovich approach presented for instance in \cite{Lifshitz_Pitaevskii_1981_Physical_Kinetics}.
The Klimontovich derivation is the more straightforward from a technical
point of view. We now recall the main steps of the Klimontovich derivation,
that will be useful later.\\

In the following we will consider statistical averages over measures
of initial conditions for the $N$ particle initial conditions $\left\{ \mathbf{r}_{n}^{0},\mathbf{v}_{n}^{0}\right\} $.
We denote $\mathbb{E}_{S}$ the average with respect to this measure
of initial conditions. As an example the measure of initial conditions
could be the product measure $\prod_{n=1}^{N}g^{0}\left(\mathbf{r}_{n}^{0},\mathbf{v}_{n}^{0}\right)\mbox{d}\mathbf{r}_{n}\mbox{d}\mathbf{v}_{n}$.
But we might consider other measures of initial conditions. We assume
that for the statistical ensemble of initial conditions, the law of
large numbers $\lim_{N\rightarrow\infty}g_{N}^{0}\left(\mathbf{r},\mathbf{v}\right)=g^{0}\left(\mathbf{r},\mathbf{v}\right)$
is valid at the initial time. This is true for instance for the product
measure. In the following, for simplicity, we restrict the discussion
to cases when the initial conditions are statistically homogenous:
$g^{0}\left(\mathbf{r},\mathbf{v}\right)=P^{0}(\mathbf{v})$. In the
following, we define $f$ as the statistical average of $g_{N}$ over
the initial conditions $f(\mathbf{v},t)=\mathbb{E}_{S}\left(g_{N}(\mathbf{r},\mathbf{v},t)\right)$.

We define the fluctuations $\delta g_{N}$ by $g_{N}(\mathbf{r},\mathbf{v},t)=f(\mathbf{v})+\delta g_{N}/\sqrt{N}$.
The scaling $1/\sqrt{N}$ is natural when we see the Vlasov equation
(\ref{eq:vlasov}) as a law of large numbers for the empirical density.
For the potential we obtain $V\left[g_{N}\right]=V\left[\delta g_{N}\right]/\sqrt{N}$,
as $f$ is homogeneous. If we introduce this decomposition in the
Klimontovich equation (\ref{eq:Klimontovich_eq}), we obtain 
\small
\begin{eqnarray}
\frac{\partial f}{\partial t} & = & \frac{1}{N}\mathbb{E}_{S}\left(\frac{\partial V\left[\delta g_{N}\right]}{\partial\mathbf{r}}.\frac{\partial\delta g_{N}}{\partial\mathbf{v}}\right)\label{eq:decomp}\\
\frac{\partial\delta g_{N}}{\partial t}+\mathbf{v}.\frac{\partial\delta g_{N}}{\partial\mathbf{r}}-\frac{\partial V\left[\delta g_{N}\right]}{\partial\mathbf{r}}.\frac{\partial f}{\partial\mathbf{v}} & = & \frac{1}{\sqrt{N}}\left[\frac{\partial V\left[\delta g_{N}\right]}{\partial\mathbf{r}}.\frac{\partial\delta g_{N}}{\partial\mathbf{v}}-\mathbb{E}_{S}\left(\frac{\partial V\left[\delta g_{N}\right]}{\partial\mathbf{r}}.\frac{\partial\delta g_{N}}{\partial\mathbf{v}}\right)\right].\label{eq:decomp2}
\end{eqnarray}
\normalsize
In the first equation, the right hand side of the equation $\frac{1}{N}\mathbb{E}_{S}\left(\frac{\partial V\left[\delta g_{N}\right]}{\partial\mathbf{r}}.\frac{\partial\delta g_{N}}{\partial\mathbf{v}}\right)$
is called the averaged non linear term and is responsible for the
long term evolution of the distribution $f$. The right hand side
of the second equation $\frac{1}{\sqrt{N}}\left[\frac{\partial V\left[\delta g_{N}\right]}{\partial\mathbf{r}}.\frac{\partial\delta g_{N}}{\partial\mathbf{v}}-\right.$ $\left.\mathbb{E}_{S}\left(\frac{\partial V\left[\delta g_{N}\right]}{\partial\mathbf{r}}.\frac{\partial\delta g_{N}}{\partial\mathbf{v}}\right)\right]$
describes the fluctuations of the non-linear term. For stable distributions
$f,$ and on timescales much smaller than $\sqrt{N}$, we can neglect
this term, following Klimontovich and classical textbooks \cite{Lifshitz_Pitaevskii_1981_Physical_Kinetics}. Please see~\cite{Caglioti_Rousset_2007_JStatPhys_QSS} for a mathematical proof of a sufficient condition of stability on time scales of order $N^\alpha$, for some $\alpha<1$.
Neglecting the terms much smaller than $\sqrt{N}$ closes the hierarchy of the correlation functions. The Bogoliubov
approximation then amounts to using the time scale separation between
the evolution of $f$ and $\delta g_{N}$. Then for fixed $f$, the
equation for $\delta g_{N}$ (\ref{eq:decomp2}) is linear when $f$
is fixed. One computes the correlation function $\mathbb{E}_{S}\left(\frac{\partial V\left[\delta g_{N}\right]}{\partial\mathbf{r}}.\frac{\partial\delta g_{N}}{\partial\mathbf{v}}\right)$
resulting from (\ref{eq:decomp2}) with fixed $f$, and argues that
this two point correlation function converges to a stationary quantity
on time scales much smaller than $\sqrt{N}$. Using this quasi-stationary
correlation function $\mathbb{E}_{S}\left(\frac{\partial V\left[\delta g_{N}\right]}{\partial\mathbf{r}}.\frac{\partial\delta g_{N}}{\partial\mathbf{v}}\right)$,
one can compute the right hand side of (\ref{eq:decomp}) as a function
of $f$. 

After time rescaling $\tau=t/N$, we define $g_{N}^{s}\left(\mathbf{r},\mathbf{v},\tau\right)=g_{N}\left(\mathbf{r},\mathbf{v},N\tau\right)$.
By abuse of notation and for convenience, we still denote $g_{N}^{s}(\tau)=g_{N}(\tau)$.
The closed equation for $g_{N}(\tau)$, which is obtained from (\ref{eq:decomp})
is the Balescu--Guernsey--Lenard equation (\ref{eq:BLG_eq}). We
do not reproduce these lengthy and classical computations that can
be found in plasma physics textbooks, for instance in Chapter
51 of \cite{Lifshitz_Pitaevskii_1981_Physical_Kinetics}. A natural
conjecture is that we have a law of large numbers $\lim_{N\rightarrow\infty}g_{N}\left(\mathbf{r},\mathbf{v},\tau\right)=f(\mathbf{v},\tau)$, where $f$ solves the Balescu--Guernsey--Lenard equation (\ref{eq:BLG_eq}), and
valid for any finite time $\tau$. 

\paragraph{Symmetries and conservation properties.}

The Balescu--Guernsey--Lenard equation (\ref{eq:BLG_eq}) has several
important physical properties:
\begin{enumerate}
\item It conserves the mass $M[f]$, momentum $\mathbf{P}[f]$ and total
kinetic energy $E[f]$ defined by
\begin{equation}
M[f]=\int\text{d}\mathbf{v}\,f\left(\mathbf{v}\right),\,\,\,\mathbf{P}[f]=\int\text{d}\mathbf{v}\,\textbf{v}f\left(\mathbf{v}\right)\,\,\,\text{and}\,\,\,E[f]=\int\text{d}\mathbf{v\,}\frac{\textbf{v}^{2}}{2}f\left(\mathbf{v}\right).\label{eq:Conservation_Laws}
\end{equation}
\item It increases monotonically the entropy $S[f]$ defined by
\[
S[f]=-k_{B}\int\text{d}\mathbf{v}\,f\left(\mathbf{v}\right)\log f\left(\mathbf{v}\right),
\]
where $k_{B}$ is the Boltzmann constant.
\item It converges towards the Boltzmann distribution for the corresponding
energy
\[
f_{B}\left(\mathbf{v}\right)=\frac{\beta{}^{3/2}}{\left(2\pi\right)^{3/2}}\exp\left(-\beta\frac{\mathbf{v}^{2}}{2}\right).
\]
\end{enumerate}

\section{Derivation of the large deviation principle from the quasi-linear
dynamics\label{sec:Derivation-of-the}}

In this section, we derive a large deviation principle for the empirical
density of $N$ particles with long range interactions, directly from
the dynamics (\ref{eq:dynamics}). 

In section \ref{subsec:The-quasi-linear-model}, we introduce the
quasi-linear dynamics of the empirical density of $N$ long range
interacting particles, for which the law of large numbers is the Balescu--Guernsey--Lenard
kinetic theory. In section \ref{subsec:The-quasi-linear-model}, we
explain that this quasi-linear dynamics for the empirical density
can be seen as a slow-fast system, for which we can define the path
large deviation functional for the slow variable. In section \ref{subsec:The-quasi-stationary-Gaussian},
we characterize the stochastic process for the quasi-linear dynamics
of the fluctuations of the empirical density as a stationary Gaussian
process. 

\subsection{The Klimontovich approach, quasilinear and slow-fast dynamics\label{subsec:The-quasi-linear-model}}

We begin by equations which are similar to (\ref{eq:decomp}-\ref{eq:decomp2}),
but by contrast to the discussion of the previous section, we will
not compute just the average for the effect of fluctuations on the
evolution of $f_{N}$, but all the cumulants after time averaging. 

We consider the empirical density
\[
g_{N}\left(\mathbf{r},\mathbf{v},t\right)=\frac{1}{N}\sum_{n=1}^{N}\delta\left(\mathbf{v}-\mathbf{v}_{n}\left(t\right)\right)\delta\left(\mathbf{r}-\mathbf{r}_{n}\left(t\right)\right),
\]
of $N$ particles which interact through a long range pair potential
according to the dynamics (\ref{eq:dynamics}). From these equations
of motion, we can deduce the Klimontovich equation
\begin{equation}
\frac{\partial g_{N}}{\partial t}+\mathbf{v}\cdot\frac{\partial g_{N}}{\partial\mathbf{r}}-\frac{\partial V\left[g_{N}\right]}{\partial\mathbf{r}}\cdot\frac{\partial g_{N}}{\partial\mathbf{v}}=0.\label{eq:Klimontovich}
\end{equation}

We consider the decomposition
\[
g_{N}\left(\mathbf{r},\mathbf{v},t\right)=f_{N}\left(\mathbf{v}\right)+\frac{1}{\sqrt{N}}\delta g_{N}\left(\mathbf{r},\mathbf{v},t\right),
\]
where $f_{N}\left(\mathbf{v},t\right)=\frac{1}{L^{3}}\int\text{d}\mathbf{r}\,g_{N}\left(\mathbf{r},\mathbf{v},t\right)$
is the projection of $g_{N}$ on homogeneous distributions (distributions
that depend on velocity only) and $\delta g_{N}$ describes the inhomogeneous
fluctuations of the empirical density $g_{N}$. Alternately, we can
understand $f_{N}$ as the empirical density of the $N$ particles
in the velocity space: $f_{N}\left(\mathbf{r},\mathbf{v},t\right)=N^{-1}L^{-3}\sum_{n=1}^{N}\delta\left(\mathbf{v}-\mathbf{v}_{n}\left(t\right)\right)$.
From the Klimontovich equation (\ref{eq:Klimontovich}), we straightforwardly
write 
\begin{eqnarray}
\frac{\partial f_{N}}{\partial t} & = & \frac{1}{NL^{3}}\int\text{d}\mathbf{r}\,\left(\frac{\partial V\left[\delta g_{N}\right]}{\partial\mathbf{r}}.\frac{\partial\delta g_{N}}{\partial\mathbf{v}}\right),\label{eq:slow}\\
\frac{\partial\delta g_{N}}{\partial t} & = & -\mathbf{v}.\frac{\partial\delta g_{N}}{\partial\mathbf{r}}+\frac{\partial V\left[\delta g_{N}\right]}{\partial\mathbf{r}}.\frac{\partial f_{N}}{\partial\mathbf{v}}\\
& &+\frac{1}{\sqrt{N}}\left[\frac{\partial V\left[\delta g_{N}\right]}{\partial\mathbf{r}}.\frac{\partial\delta g_{N}}{\partial\mathbf{v}}-\frac{1}{L^{3}}\int\text{d}\mathbf{r}\,\left(\frac{\partial V\left[\delta g_{N}\right]}{\partial\mathbf{r}}.\frac{\partial\delta g_{N}}{\partial\mathbf{v}}\right)\right].\label{eq:fast}
\end{eqnarray}
Just like in the previous section, we will consider statistical averages over a probability measure for the initial conditions $\left\{ \mathbf{r}_{n}^{0},\mathbf{v}_{n}^{0}\right\}$ of the $N$ particles. As the microscopic dynamics is deterministic, the only source of randomness is the ensemble of initial conditions. We assume that this ensemble of initial conditions is sampled from a spatially homogeneous measure and that the set of corresponding $g_N$ is concentrated close to homogeneous distributions in the $(\mathbf{r},\mathbf{v})$ space. Moreover we assume that the large deviation principle 
\begin{equation}
\mathbf{P}\left( f_{N}\left(\tau=0\right) = f^0 \right) \underset{N\rightarrow\infty}{\asymp} \text{e}^{-NI_{0}\left[f^0\right]},
\label{eq:I_0}
\end{equation}
holds, where $I_{0}$ is a large deviation rate function for $f_{N}(\tau=0)$
the initial conditions of $f_{N}$. As an example, the measure of initial conditions $\left\{ \mathbf{r}_{n}^{0},\mathbf{v}_{n}^{0}\right\}$
could be the homogeneous product measure $\prod_{n=1}^{N}P^0\left(\mathbf{v}_{n}^{0}\right)\mbox{d}\mathbf{v}_{n}\mbox{d}\mathbf{r}_{n}/L^3$, with $\int \mbox{d}\mathbf{v} P^0\left(\mathbf{v}\right) =1$. Then $I_{0}$ would then be the Kullback--Leibler divergence of $f^0$ with respect to $P^0$. But we might consider other ensembles of initial conditions. 

We now assume the validity of the quasi-linear approximation, which
amounts to neglecting terms of order $N^{-1/2}$ in the evolution equation
for $\delta g_{N}$. We also change the timescale $\tau=t/N$ and
obtain the quasilinear dynamics

\begin{eqnarray}
\frac{\partial f_{N}}{\partial\tau} & = & \frac{1}{L^{3}}\int\text{d}\mathbf{r}\,\left(\frac{\partial V\left[\delta g_{N}\right]}{\partial\mathbf{r}}.\frac{\partial\delta g_{N}}{\partial\mathbf{v}}\right),\label{eq:slow-1}\\
\frac{\partial\delta g_{N}}{\partial\tau} & =N & \left\{ -\mathbf{v}.\frac{\partial\delta g_{N}}{\partial\mathbf{r}}+\frac{\partial V\left[\delta g_{N}\right]}{\partial\mathbf{r}}.\frac{\partial f_{N}}{\partial\mathbf{v}}\right\} .\label{eq:fast-1}
\end{eqnarray}
When $N$ goes to infinity, we observe that the equation for $\delta g_{N}$
is a fast process, with timescales for $\tau$ of order $1/N$, while
the equation for $f_{N}$ is a slow one with timescales for $\tau$
of order $1$. For such slow-fast dynamics, it is natural to consider
$f_{N}$ fixed (frozen) in equation (\ref{eq:fast-1}) on time scales
for $\tau$ of order $1/N.$ For fixed $f_{N}$ the dynamics for $\delta g_{N}$
is linear and can be solved. Computing then the average of the term
$\int\text{d}\mathbf{r}\,\frac{\partial V\left[\delta g_{N}\right]}{\partial\mathbf{r}}.\frac{\partial\delta g_{N}}{\partial\mathbf{v}}$,
for the asymptotic process for $\delta g_{N}$ for fixed $f_{N}$
leads to the Guernsey--Lenard--Balescu equation, as explained in
section 3. Those computation can be found in classical textbooks \cite{Lifshitz_Pitaevskii_1981_Physical_Kinetics}.\\

In the following we want to go beyond these classical computations,
by estimating not just the average of the left hand side in (\ref{eq:slow}),
$\int\text{d}\mathbf{r}\,\frac{\partial V\left[\delta g_{N}\right]}{\partial\mathbf{r}}.\frac{\partial\delta g_{N}}{\partial\mathbf{v}}$,
but all the cumulants of the time averages $\int_{0}^{\Delta T}\int\text{d}\mathbf{r}\,\frac{\partial V\left[\delta g_{N}\right]}{\partial\mathbf{r}}.\frac{\partial\delta g_{N}}{\partial\mathbf{v}}$
in order to describe the large deviations for the process $f_{N}$.
For slow-fast dynamics, the theory for the large deviations of the
effective evolution of the slow variable is a classical one both in
theoretical physics (see for instance \cite{Bouchet_Grafke_Tangarife_Vanden-Eijnden_2015_largedeviations})
and mathematics. In the mathematics literature, it is for instance
treated for diffusions \cite{Freidlin_Wentzel_1984_book,Veretennikov},
or chaotic deterministic systems \cite{kifer1992averaging,kifer2004averaging}.
The result for the path large deviations for the slow dynamics is
explained in section 2.2.2 of \cite{2021arXiv210104455F}
(see equations (20-21)). After rescaling time $\tau = t/N$, we then have 
\begin{equation}
\mathbf{P}\left(\left\{ f_{N}\left(\mathbf{v},\tau\right)\right\} _{0\leq\tau\leq T}=\left\{ f\left(\mathbf{v},\tau\right)\right\} _{0\leq\tau\leq T}\right)\underset{N\rightarrow\infty}{\asymp}\text{e}^{-NL^{3}\text{Sup}_{p}\int_{0}^{T}\text{d}\tau\left\{ \int\text{d}\mathbf{v}\,\dot{f}p-H[f,p]\right\} }\text{e}^{-NI_{0}\left[f_0\right]},\label{eq:PGD_H_GE}
\end{equation}
where $I_{0}$ is a large deviation rate function for 
the initial conditions of $f_{N}$, see equation (\ref{eq:I_0}), and with 
\begin{equation}
H\left[f,p\right]=\underset{T\rightarrow\infty}{\lim}\frac{1}{TL^{3}}\log\mathbb{E}_{f}\left[\text{exp\ensuremath{\left(\int_{0}^{T}\text{d}t\,\int\text{d}\mathbf{v}\,p\left(\mathbf{v}\right)\int\text{d}\mathbf{r}'\frac{\partial V\left[\delta g_{N}\right]}{\partial\mathbf{r}'}.\frac{\partial\delta g_{N}}{\partial\mathbf{v}}\right)}}\right]\label{eq:H_Gartner_ellis}
\end{equation}
and where $\mathbb{E}_{f}$ denotes the expectation on the process
for $\delta g_{N},$ evolving according to 
\begin{equation}
\frac{\partial\delta g_{N}}{\partial t}=-\mathbf{v}.\frac{\partial\delta g_{N}}{\partial\mathbf{r}}+\frac{\partial V\left[\delta g_{N}\right]}{\partial\mathbf{r}}.\frac{\partial f}{\partial\mathbf{v}}.\label{eq:Linear_deltagN_fast}
\end{equation}
In this equation, $f_{N}=f$ is fixed and time independent. We note
that the classical mathematical results to justify (\ref{eq:H_Gartner_ellis})
would require to prove mixing properties for the fast process, and
stability of the invariant measure, that nobody has proven yet for
(\ref{eq:Linear_deltagN_fast}).

We note that to obtain equation (\ref{eq:H_Gartner_ellis}) from equation
(21) of \cite{2021arXiv210104455F}, we have considered $f_{N}$ as
a function of the $\mu$-space. Then the conjugated momentum $p\left(\mathbf{r},\mathbf{v}\right)$
should also be a function of the $\mu$-space and the scalar product
be the one of the $\mu$-space. However, recognizing that for homogeneous
$f$, $p$ should also be homogeneous ($p\left(\mathbf{r},\mathbf{v}\right)=p\left(\mathbf{v}\right)$),
and performing trivial integration over $\mathbf{r}$ leads to (\ref{eq:H_Gartner_ellis}).
The $L^{3}$ factor in the large deviation principle (\ref{eq:PGD_H_GE})
also comes from a trivial integration over $\mathbf{r}$ of $\dot{f}p$.
In the definition of $H$, in (\ref{eq:H_Gartner_ellis}) we have
divided the scaled cumulant generating function by $L^{3}$ for convenience,
such that the action in (\ref{eq:PGD_H_GE}) appears as a natural
action for homogenous distributions. 

The goal of the following sections and the contribution of this work is to obtain an explicit expression
for (\ref{eq:H_Gartner_ellis}).
\subsection{The quasi-stationary Gaussian process for $\delta g_{N}$\label{subsec:The-quasi-stationary-Gaussian}}

In order to compute (\ref{eq:H_Gartner_ellis}), we need
to estimate averages over the stochastic process which corresponds
to generic sets of initial condition for $\delta g_{N}$, and where
$\delta g_{N}$ satisfies equation (\ref{eq:Linear_deltagN_fast}). We first note that for
fixed $f$, equation (\ref{eq:Linear_deltagN_fast}) is linear. If the set of initial conditions
for $\delta g_{N}(t=0)$ is a Gaussian random variable, then the stochastic
process $\left\{ \delta g_{N}(t)\right\} _{t\geq0}$ will be a Gaussian
process. Several recent mathematical works~\cite{lancellotti2009fluctuations,lancellotti2010vlasov,lancellotti2016time,paul2019size,velazquez2018two,duerinckx2021size}
discuss some properties of the Gaussian process of the fluctuations
close to a Vlasov solution. For instance~\cite{lancellotti2016time}
proves that, when starting from sets of Gaussian initial conditions
of the form of relevant central limit theorems, at long times, the stochastic
process converges to a statistically {\it stationary} Gaussian process. The
fact that for generic sets of initial conditions, the stochastic process
of the fluctuations $\delta g_{N}$ converges to a stochastically stationary process,
which is independent of the initial conditions, has long been understood by physicists.
This is for instance explained in §51 of~\cite{Lifshitz_Pitaevskii_1981_Physical_Kinetics}, where
the asymptotic stationary process is precisely characterized. This striking convergence result is related
to the Landau damping and the fact that we deal with particle systems.
The work~\cite{Bouchet:2004_PRE_StochasticProcess} derives another characterization of this stationary
process, based on an integral equation, and illustrates numerically
the convergence. In the following we will thus consider averages in
equation (\ref{eq:H_Gartner_ellis}) as averages
over this stationary Gaussian process. Such stationary averages
are denoted $\mathbb{E}_{S}$.

We do not reproduce the classical and lengthy computations of the correlation
functions of this stationary process, but just report the formulas
which can be found for instance in §51 of \cite{Lifshitz_Pitaevskii_1981_Physical_Kinetics}.
The potential autocorrelation function are homogeneous because of
the space translation symmetry. Then 
\[
\mathbb{E}_{S}\left(V\left[\delta g_{N}\right]\left(\mathbf{r}_{1},t_{1}\right)V\left[\delta g_{N}\right]\left(\mathbf{r}_{2},t_{2}\right)\right)=\mathcal{C}_{VV}\left(\mathbf{r}_{1}-\mathbf{r}_{2},t_{1}-t_{2}\right),
\]
We define $\tilde{\varphi}$ the space-time Fourier transform of a
function $\varphi$ as 
\begin{equation}
\tilde{\varphi}\left(\mathbf{k},\omega\right)=\int_{\left[0,L\right]^{3}}\text{d}\mathbf{r}\int_{-\infty}^{\infty}\text{d}t\,\text{e}^{-i\left(\mathbf{k}.\mathbf{r}-\omega t\right)}\varphi\left(\mathbf{r},t\right),\label{eq:FL_Transf}
\end{equation}
following the same convention as in \cite{Lifshitz_Pitaevskii_1981_Physical_Kinetics}.
According to equation (51.20), §51 of \cite{Lifshitz_Pitaevskii_1981_Physical_Kinetics},
with the identification $V=e\phi$ and $\hat{W}(k)=4\pi e^{2}/k^{2}$,
the space-time Fourier transform of the autocorrelation function of
the potential then reads
\begin{equation}
\widetilde{\mathcal{C}_{VV}}\left(\mathbf{k},\omega\right)=2\pi\left[\int\text{d}\mathbf{v}'\,f\left(\mathbf{v}'\right)\delta\left(\omega-\mathbf{k}.\mathbf{v}'\right)\right]\frac{\hat{W}\mathbf{\left(k\right)}^{2}}{\left|\varepsilon\left[f\right]\left(\mathbf{k},\omega\right)\right|^{2}}.\label{eq:Autocorrelation_Potentiel_Fourier_Laplace}
\end{equation}

Similarly the time stationary correlation function between the potential
and distribution fluctuation is space-time homogeneous 
\[ \mathbb{E}_{S}\left(V\left[\delta g_{N}\right]\left(\mathbf{r}_{1},t_{1}\right)\delta g_{N}\left(\mathbf{r}_{2},\text{\ensuremath{\mathbf{v}}},t_{2}\right)\right)=\mathcal{C}_{VG}\left(\mathbf{r}_{1}-\mathbf{r}_{2},t_{1}-t_{2},\mathbf{v}\right).
\]
According to equation (51.21) of \cite{Lifshitz_Pitaevskii_1981_Physical_Kinetics},
its space-time Fourier transform reads
\begin{equation}
\widetilde{\mathcal{C}_{VG}}\left(\mathbf{k},\omega,\mathbf{v}\right)=-\frac{\mathbf{k}}{\omega-\mathbf{k}.\mathbf{v}-i\eta}.\frac{\partial f}{\partial\mathbf{v}}\left(\mathbf{v}\right)\widetilde{\mathcal{C}_{VV}}\left(\mathbf{k},\omega\right)+2\pi\frac{\hat{W}\left(\mathbf{k}\right)}{\varepsilon\left[f\right]\left(\mathbf{k},\omega\right)}f\left(\mathbf{v}\right)\delta\left(\omega-\mathbf{k}.\mathbf{v}\right).\label{eq:Autocorrelation_VG_Fourier_Laplace}
\end{equation}
We also define the autocorrelation function of the distribution fluctuations
\[
\mathbb{E}_{S}\left(\delta g_{N}\left(\mathbf{r}_{1},\mathbf{v}_{1},t_{1}\right)\delta g_{N}\left(\mathbf{r}_{2},\mathbf{v}_{2},t_{2}\right)\right)=\mathcal{C}_{GG}\left(\mathbf{r}_{1}-\mathbf{r}_{2},t_{1}-t_{2},\mathbf{v}_{1},\mathbf{v}_{2}\right).
\]
According to equation (51.23) of \cite{Lifshitz_Pitaevskii_1981_Physical_Kinetics},
its space-time Fourier transform reads
\begin{eqnarray}
\widetilde{\mathcal{C}_{GG}}\left(\mathbf{k},\omega,\mathbf{v}_{1},\mathbf{v}_{2}\right) & = & 2\pi\delta\left(\mathbf{v}_{1}-\mathbf{v}_{2}\right)f\left(\mathbf{v}_{1}\right)\delta\left(\omega-\mathbf{k}.\mathbf{v}_{1}\right)\label{eq:Autocorrelation_GG_Fourier_Laplace}\\
 & + & \frac{\widetilde{\mathcal{C}_{VV}}\left(\mathbf{k},\omega\right)}{\left(\omega-\mathbf{k}.\mathbf{v}_{1}+i\eta\right)\left(\omega-\mathbf{k}.\mathbf{v}_{2}-i\eta\right)}\mathbf{k}.\frac{\partial f}{\partial\mathbf{v}}\left(\mathbf{v}_{1}\right)\mathbf{k}.\frac{\partial f}{\partial\mathbf{v}}\left(\mathbf{v}_{2}\right)\nonumber \\
 & - & 2\pi\hat{W}\mathbf{\left(k\right)}\mathbf{k}.\frac{\partial f}{\partial\mathbf{v}}\left(\mathbf{v}_{1}\right)\frac{f(\mathbf{v}_{2})\delta\left(\omega-\mathbf{k}.\mathbf{v}_{2}\right)}{\varepsilon\left(\mathbf{k},\omega\right)\left(\omega-\mathbf{k}.\mathbf{v}_{1}+i\eta\right)}\nonumber \\
 & - & 2\pi\hat{W}\mathbf{\left(k\right)}\mathbf{k}.\frac{\partial f}{\partial\mathbf{v}}\left(\mathbf{v}_{2}\right)\frac{f(\mathbf{v}_{1})\delta\left(\omega-\mathbf{k}.\mathbf{v}_{1}\right)}{\varepsilon^{*}\left(\mathbf{k},\omega\right)\left(\omega-\mathbf{k}.\mathbf{v}_{2}-i\eta\right)}.\nonumber 
\end{eqnarray}
We note that the order in the correlation functions for $V$ and $g_{N}$
matters. We have 
\[
\mathbb{E}_{S}\left(\delta g_{N}\left(\mathbf{r}_{1},\text{\ensuremath{\mathbf{v}}},t_{1}\right)V\left[\delta g_{N}\right]\left(\mathbf{r}_{2},t_{2}\right)\right)=\mathcal{C}_{GV}\left(\mathbf{r}_{1}-\mathbf{r}_{2},t_{1}-t_{2},\mathbf{v}\right),
\]
with 
\[\widetilde{\mathcal{C}_{VG}}\left(\mathbf{k},\omega,\mathbf{v}\right)=\widetilde{\mathcal{C}_{GV}}\left(-\mathbf{k},-\omega,\mathbf{v}\right)=\widetilde{\mathcal{C}_{GV}}^{*}\left(\mathbf{k},\omega,\mathbf{v}\right).
\]
We also note the symmetry property for $\widetilde{\mathcal{C}_{GG}}$:
$\widetilde{\mathcal{C}_{GG}}\left(\mathbf{k},\omega,\mathbf{v}_{1},\mathbf{v}_{2}\right)=\widetilde{\mathcal{C}_{GG}}\left(-\mathbf{k},-\omega,\mathbf{v}_{2},\mathbf{v}_{1}\right)$.
It is a consequence of the symmetry $\mathcal{C}_{GG}\left(\mathbf{r},t,\mathbf{v}_{1},\mathbf{v}_{2}\right)=\mathcal{C}_{GG}\left(-\mathbf{r},-t,\mathbf{v}_{2},\mathbf{v}_{1}\right)$.
Moreover, since $C_{GG}$ is real, we have $\widetilde{\mathcal{C}_{GG}}\left(-\mathbf{k},-\omega,\mathbf{v}_{2},\mathbf{v}_{1}\right)=\widetilde{\mathcal{C}_{GG}}^{*}\left(\mathbf{k},\omega,\mathbf{v}_{2},\mathbf{v}_{1}\right)$.
We thus have the symmetry 
\begin{equation}
\widetilde{\mathcal{C}_{GG}}\left(\mathbf{k},\omega,\mathbf{v}_{1},\mathbf{v}_{2}\right)=\widetilde{\mathcal{C}_{GG}}^{*}\left(\mathbf{k},\omega,\mathbf{v}_{2},\mathbf{v}_{1}\right).\label{eq:Symetrie_CGG}
\end{equation}
We note that as a mere consequence of the definition of $V\left[\delta g_{N}\right]$,
we have the following relations between the two-point correlation
functions
\begin{gather}
\widetilde{\mathcal{C}_{VG}}\left(\mathbf{k},\omega,\mathbf{v}_{1}\right)=\hat{W}\mathbf{\left(k\right)}\int\text{d}\mathbf{v}_{2}\,\widetilde{\mathcal{C}_{GG}}\left(\mathbf{k},\omega,\mathbf{v}_{1},\mathbf{v}_{2}\right),\label{eq:relations_C_1}\\
\widetilde{\mathcal{C}_{VV}}\left(\mathbf{k},\omega\right)=\left(\hat{W}\mathbf{\left(k\right)}\right)^{2}\int\text{d}\mathbf{v}_{1}\text{d}\mathbf{v}_{2}\,\widetilde{\mathcal{C}_{GG}}\left(\mathbf{k},\omega,\mathbf{v}_{1},\mathbf{v}_{2}\right).\label{eq:relations_C_2}
\end{gather}

\section{Computation of the large deviation Hamiltonian\label{sec:Computation-of-the}}

In this section, we obtain an explicit formula for the large deviation
functional of the empirical density of $N$ particles with long range
interactions, starting from equation (\ref{eq:H_Gartner_ellis}).
We noticed in section \ref{subsec:The-quasi-stationary-Gaussian}
that the fluctuations of the homegeneous part empirical density are described by the average of a quadratic form over a
Gaussian stationary process. In section \ref{subsec:The-large-deviation},
we explain how this makes the computation of the Hamiltonian (\ref{eq:H_Gartner_ellis})
equivalent to the computation of a functional determinant. In section
\ref{subsec:Application-of-the}, this functional determinant is explicitly
computed, using the Szegö--Widom theorem and an explicit computation
of determinants in the space of observable over velocity distributions.

\subsection{The large deviation Hamiltonian as a functional Gaussian integral
\label{subsec:The-large-deviation}}

Within the quasi-linear approximation, the fluctuations of the empirical
density $\delta g_{N}$ describe a stationary Gaussian process over
functions of the $\mu$-space. The goal of this subsection is to show
that the computation of the large deviation Hamiltonian is equivalent
to the computation of a Gaussian functional integral of the fluctuations
of the empirical density $\delta g_{N}$.

We consider  $\mathcal{H}_{\mathbf{v}}$, the Hilbert space of complex functions over the velocity space, with $\left\langle .,.\right\rangle $, the Hermitian product:
$\left\langle a,b\right\rangle =\int\text{d}\mathbf{v}\,a^{*}\left(\mathbf{v}\right)b\left(\mathbf{v}\right)$.  We can conveniently express the argument of the exponential in the formula (\ref{eq:H_Gartner_ellis})
for the large deviation Hamiltonian using a spatial Fourier decomposition
of the fluctuations of the empirical density $\delta\hat{g}\left(\mathbf{k},\mathbf{v},t\right)=\int_{\left[0,L\right]^{3}}\text{d}\mathbf{r}\,\text{e}^{-i\mathbf{k}.\mathbf{r}}\delta g_{N}\left(\mathbf{r},\mathbf{v},t\right)$. Using this Fourier decomposition, the definition of the potential $V \left[ \delta g_N \right]$ and partial integration with respect to the velocity integral, we obtain

\begin{equation}
\ensuremath{\int\text{d}\mathbf{v}\,p\left(\mathbf{v}\right)\int\text{d}\mathbf{r}'\frac{\partial V\left[\delta g_{N}\right]}{\partial\mathbf{r}'}.\frac{\partial\delta g_{N}}{\partial\mathbf{v}}}=\frac{1}{2}\sum_{\mathbf{k}\in\left(2\pi/L\right)\mathbb{Z}^{3}}\langle \delta\hat{g}_{N}\left(\mathbf{k},\cdot,t\right),\mathbf{M}\left(\mathbf{k}\right)\left[\delta\hat{g}_{N}\left(\mathbf{k},\cdot,t\right)\right]\rangle, \label{eq:fourier-decomp}
\end{equation}
where we define the Hermitian operator $\mathbf{M}\left(\mathbf{k}\right)$
acting on $\varphi \in \mathcal{H}_{\mathbf{v}}$ as
\begin{equation}
\mathbf{M}\left(\mathbf{k}\right)\left[\varphi\right]\left(\mathbf{v}_{1}\right)=\int\text{d}\mathbf{v}_{2}\,M\left(\mathbf{k};\mathbf{v}_{1},\mathbf{v}_{2}\right)\varphi\left(\mathbf{v}_{2}\right),\label{eq:operator-m}
\end{equation}
with the kernel $M$ defined by
\begin{equation}
M\left(\mathbf{k};\mathbf{v}_{1},\mathbf{v}_{2}\right)=\frac{i}{L^{6}}\hat{W}\left(\mathbf{k}\right)\mathbf{k}.\left\{ -\frac{\partial p}{\partial\mathbf{v}}\left(\mathbf{v}_{1}\right)+\frac{\partial p}{\partial\mathbf{v}}\left(\mathbf{v}_{2}\right)\right\}.\label{eq:kernel-m}
\end{equation}
There is a factor $1/2$ on the r.h.s. of \eqref{eq:fourier-decomp} because we chose to  symmetrize the expression, such that $M\left(\mathbf{k};\mathbf{v}_{1},\mathbf{v}_{2}\right)=M\left(\mathbf{k};\mathbf{v}_{2},\mathbf{v}_{1}\right)^*$.  $\mathbf{M}\left(\mathbf{k}\right)$ is then an Hermitian operator. 

The goal of the following of this subsection is to express the sum on the r.h.s. of \eqref{eq:fourier-decomp} as a sum of independent terms to make the computation of \eqref{eq:H_Gartner_ellis} easier. Since  for every $\mathbf{k}\in\left(2\pi/L\right)\mathbb{Z}^{3}$, $\mathbf{M}\left(\mathbf{k}\right)$ is an Hermitian operator, $\mathbf{M}\left(\mathbf{k}\right)^*=\mathbf{M}\left(-\mathbf{k}\right)$, and
\[
\delta\hat{g}_{N}^*\left(\mathbf{k},\cdot,t\right)=\delta\hat{g}_{N}\left(\mathbf{-k},\cdot,t\right),
\] 
we have the following relation
\begin{equation}
\langle \delta\hat{g}_{N}\left(\mathbf{k},\cdot,t\right),\mathbf{M}\left(\mathbf{k}\right)\left[\delta\hat{g}_{N}\left(\mathbf{k},\cdot,t\right)\right]\rangle =\langle \delta\hat{g}_{N}\left(-\mathbf{k},\cdot,t\right),\mathbf{M}\left(\mathbf{-k}\right)\left[\delta\hat{g}_{N}\left(\mathbf{-k},\cdot,t\right)\right]\rangle .\label{eq:k,-k,symmetry}
\end{equation}
This implies that on the r.h.s. of \eqref{eq:fourier-decomp},
the contribution of an index $\mathbf{k}\in\left(2\pi/L\right)\mathbb{Z}^{3}$
will be equal to the contribution of its negative $-\mathbf{k}$.

Because the stochastic process $\delta g_N (\mathbf{r},\cdot,t)$, for the fluctuations of the distribution function is spatially homogeneous, the stochastic process $\delta\hat{g}_{N}\left(\mathbf{k},\cdot,t\right)$
is statistically mutually independent with every other $\delta\hat{g}_{N}\left(\mathbf{k}',\cdot,t\right)$
as long as $\mathbf{k}'\neq-\mathbf{k}$. Because $\delta\hat{g}_{N}\left(\mathbf{k},\cdot,t\right)$ is not statistically independent from $\delta\hat{g}_{N}\left(-\mathbf{k},\cdot,t\right)$, it is useful to treat them together. 
We define $\mathbb{Z}_{\pm}^{3}=\mathbb{Z}^{3}/Z_{2}$ the quotient of group $\mathbb{Z}^{3}$ with $Z_{2}$, the cyclic group of order 2. In
other words, $\mathbb{Z}_{\pm}^{3}$ is the set of triplets of integers
where we identify a triplet $(a,b,c)\in\mathbb{Z}^{3}$ with its negative
$(-a,-b,-c)$. Then, using (\ref{eq:k,-k,symmetry}), the sum over $\mathbf{k}\in\left(2\pi/L\right)\mathbb{Z}^{3}$
can be rewritten as 
\small
\begin{equation}
\frac{1}{2}\sum_{\mathbf{k}\in\left(2\pi/L\right)\mathbb{Z}^{3}}\left\langle \delta\hat{g}_{N}\left(\mathbf{k},\cdot,t\right),\mathbf{M}\left(\mathbf{k}\right)\left[\delta\hat{g}_{N}\left(\mathbf{k},\cdot,t\right)\right]\right\rangle =\sum_{\mathbf{k}\in\left(2\pi/L\right)\mathbb{Z}_{\pm}^{3}}\left\langle \delta\hat{g}_{N}\left(\mathbf{k},\cdot,t\right),\mathbf{M}\left(\mathbf{k}\right)\left[\delta\hat{g}_{N}\left(\mathbf{k},\cdot,t\right)\right]\right\rangle .\label{eq:Z2}
\end{equation}
\normalsize
As a consequence, the r.h.s
of (\ref{eq:Z2}) is a sum of statistically independent terms. We can then use the fact that the expected value of a product of independent random variables is the product of their expected values, as well as equations \eqref{eq:fourier-decomp} and \eqref{eq:Z2} to obtain
\small
\begin{equation}
\mathbb{E}\left[\text{exp\ensuremath{\left(\int_{0}^{T}\text{d}t\,\int\text{d}\mathbf{v}\,p\left(\mathbf{v}\right)\int\text{d}\mathbf{r}'\frac{\partial V\left[\delta g_{N}\right]}{\partial\mathbf{r}'}.\frac{\partial\delta g_{N}}{\partial\mathbf{v}}\right)}}\right]=\prod_{\mathbf{k}\in\left(2\pi/L\right)\mathbb{Z}_{\pm}^{3}}\mathbb{E}\left[\exp\left(\int_{0}^{T}\text{d}t\,\langle \delta\hat{g}_{N}\left(\mathbf{k},\cdot,t\right),\mathbf{M}\left(\mathbf{k}\right)\left[\delta\hat{g}_{N}\left(\mathbf{k},\cdot,t\right)\right]\rangle \right)\right].
\label{eq:expectation}
\end{equation}
\normalsize

We can then go back to \eqref{eq:H_Gartner_ellis} using \eqref{eq:k,-k,symmetry} and \eqref{eq:expectation} to express the large deviation Hamiltonian as a sum over the wavevectors
\begin{equation}
H\left[f,p\right]=\sum_{\mathbf{k}\in\left(2\pi/L\right)\mathbb{Z}^{3}}\hat{H}\left[f,p\right]\left(\mathbf{k}\right),\label{eq:H-decomp-1}
\end{equation}
where
\begin{equation}
\hat{H}\left[f,p\right]\left(\mathbf{k}\right)=\underset{T\rightarrow\infty}{\lim}\frac{1}{2TL^{3}}\log\mathbb{E}\left[\text{exp\ensuremath{\left(\int_{0}^{T}\text{d}t\,\left\langle \delta\hat{g}_{N}\left(\mathbf{k},\cdot,t\right),\mathbf{M}\left(\mathbf{k}\right)\left[\delta\hat{g}_{N}\left(\mathbf{k},\cdot,t\right)\right]\right\rangle \right)}}\right].\label{eq:H-hat}
\end{equation}

\subsection{Application of the Szegö--Widom theorem \label{subsec:Application-of-the}}

The computation of (\ref{eq:H-decomp-1}-\ref{eq:H-hat}) requires
to estimate large time large deviations of a quadratic functional
of a Gaussian stochastic process. More precisely, the Gaussian process
involved in (\ref{eq:H-hat}) is the stochastic
process of the $\mathbf{k}$-th Fourier mode of the fluctuations of
the empirical density $\delta\hat{g}_{N}\left(\mathbf{k},\cdot,t\right)$,
and the quadratic functional is defined by the Hermitian operator
$\mathbf{M}\left(\mathbf{k}\right)$ (\ref{eq:operator-m}). Since
$\delta\hat{g}_{N}\left(\mathbf{k},\cdot,t\right)$ is a Gaussian
process, it is possible to compute (\ref{eq:H-hat})
via functional determinants. Thanks to the Szegö--Widom theorem, it is possible
to evaluate the asymptotics of this Fredholm determinant in terms
of much simpler determinants of an operator on $\mathcal{H}_{\mathbf{v}}$.
This program was first implemented in \cite{Bouchet-Roger-Zaboronski}, with a nice application to a model inspired by 2D and geophysical turbulence. In appendix \ref{sec:Long-time-large} we explain the details of this program, easily adapting \cite{Bouchet-Roger-Zaboronski}, for Gaussian processes with complex variables. The result (\ref{eq:asymptotics_quadratic_observable}) of the appendix
\ref{sec:Long-time-large}, adapted to the case where the Hilbert
space is $\mathcal{H}_{\mathbf{v}}$, reads
\begin{equation}
\log\mathbb{E}\left[\text{exp\ensuremath{\left(\int_{0}^{T}\text{d}t\left\langle \delta\hat{g}_{N}\left(\mathbf{k},\cdot,t\right),\mathbf{M}\left(\mathbf{k}\right)\left[\delta\hat{g}_{N}\left(\mathbf{k},\cdot,t\right)\right]\right\rangle \right)}}\right]\underset{T\rightarrow\infty}{\sim}-\frac{T}{2\pi}\int\text{d}\omega\,\log\underset{\mathcal{H}_{\mathbf{v}}}{\det}\left(u_{\mathbf{k},\omega}\right),\label{eq:asymptotics_quadratic_observable-1}
\end{equation}
where, for any $\mathbf{k}$ and $\omega$, and $\varphi\in\mathcal{H}_{\mathbf{v}}$,
$u_{\mathbf{k,\omega}}\left[\varphi\right]$ is defined by
\begin{eqnarray*}
u_{\mathbf{k},\omega}\left[\varphi\right]\left(\mathbf{v}_{1}\right) & = & \varphi\left(\mathbf{v}_{1}\right)+\int\text{d}\mathbf{v}_{2}\text{d}\mathbf{v}_{3}\,M\left(\mathbf{k};\mathbf{v}_{1},\mathbf{v}_{2}\right)\widetilde{\mathcal{C}_{GG}}\left(\mathbf{k},\omega,\mathbf{v}_{2},\mathbf{v}_{3}\right)\varphi\left(\mathbf{v}_{3}\right).
\end{eqnarray*}
$u_{\mathbf{k},\omega}$ is a linear operator of $\mathcal{H}_{\mathbf{v}}$.
The subscript $\mathcal{H}_{\mathbf{v}}$ in (\ref{eq:asymptotics_quadratic_observable-1})
indicates that the determinant is a determinant of an operator over
$\mathcal{H}_{\mathbf{v}}$. 

Then, combining equations (\ref{eq:H-hat})
and (\ref{eq:asymptotics_quadratic_observable-1}) yields
\begin{equation}
\hat{H}\left[f,p\right]\left(\mathbf{k}\right)=-\frac{1}{4\pi L^{3}}\int\text{d}\omega\,\log\underset{\mathcal{H}_{\mathbf{v}}}{\det}\left(u_{\mathbf{k},\omega}\right).\label{eq:H_det}
\end{equation}\\

Our next task is to obtain an explicit formula for
$\hat{H}\left[f,p\right]\left(\mathbf{k}\right)$ and thus for the
full large deviation Hamiltonian $H$ (\ref{eq:H-decomp-1}) is to compute
$\underset{\mathcal{H}_{\mathbf{v}}}{\det}\left(u_{\mathbf{k},\omega}\right)$.
This determinant can be easily computed once we realize that the range
of $u_{\mathbf{k},\omega}-\text{Id}$ is two-dimensional. The explicit
computation is performed in appendix \ref{sec:determinant}. The result
reads 
\begin{equation}
\underset{\mathcal{H}_{\mathbf{v}}}{\det}\left(u_{\mathbf{k},\omega}\right)=1-\mathcal{J}\left[f,p\right]\left(\mathbf{k},\omega\right),\label{eq:det_u}
\end{equation}
with
\begin{multline}
\mathcal{J}\left[f,p\right]\left(\mathbf{k},\omega\right)=-2\int\text{d}\mathbf{v}_{1}\,\mathbf{k}.\frac{\partial p}{\partial\mathbf{v}_{1}}\Im\left(\widetilde{\mathcal{C}_{VG}}\left(\mathbf{k},\omega,\mathbf{v}_{1}\right)\right)\\
-\int\text{d}\mathbf{v}_{1}\text{d}\mathbf{v}_{2}\,\mathbf{k}.\frac{\partial p}{\partial\mathbf{v}_{1}}\mathbf{k}.\frac{\partial p}{\partial\mathbf{v}_{2}}\left\{ \widetilde{\mathcal{C}_{VG}}\left(\mathbf{k},\omega,\mathbf{v}_{1}\right)\widetilde{\mathcal{C}_{VG}}\left(\mathbf{k},\omega,\mathbf{v}_{2}\right)^{*}\right. \\
- \left. \widetilde{\mathcal{C}_{VV}}\left(\mathbf{k},\omega\right)\widetilde{\mathcal{C}_{GG}}\left(\mathbf{k},\omega,\mathbf{v}_{1},\mathbf{v}_{2}\right)\right\} .\label{eq:det_general}
\end{multline}
Using the expressions of the two-point correlation functions (\ref{eq:Autocorrelation_Potentiel_Fourier_Laplace}-\ref{eq:Autocorrelation_GG_Fourier_Laplace}),
we obtain that 
\begin{equation}
\mathcal{J}\left[f,p\right]=\mathcal{L}\left[f,p\right]+Q\left[f,p,p\right],\label{eq:J}
\end{equation}
where $\mathcal{L}$ depends linearly on $p$ and $Q$ depends on
$p$ as a quadratic form. We have 
\begin{equation}
\mathcal{L}\left[f,p\right]\left(\mathbf{k},\omega\right)=4\pi\int\text{d}\mathbf{v}_{1}\text{d}\mathbf{v}_{2}\,\mathbf{A}\left[f\right]\left(\mathbf{k},\omega,\mathbf{v}_{1},\mathbf{v}_{2}\right):\frac{\partial p}{\partial\mathbf{v}_{1}}\left\{ \frac{\partial f}{\partial\mathbf{v}_{2}}f(\mathbf{v}_{1})-f(\mathbf{v}_{2})\frac{\partial f}{\partial\mathbf{v}_{1}}\right\} \label{eq:L}
\end{equation}
and 
\begin{multline}
Q\left[f,p,q\right]\left(\mathbf{k},\omega\right)=2\pi\int\text{d}\mathbf{v}_{1}\text{d}\mathbf{v}_{2}\,\mathbf{A}\left[f\right]\left(\mathbf{k},\omega,\mathbf{v}_{1},\mathbf{v}_{2}\right):\left\{ \frac{\partial p}{\partial\mathbf{v}_{1}}\frac{\partial q}{\partial\mathbf{v}_{1}}\right. \\
+\left. \frac{\partial p}{\partial\mathbf{v}_{2}}\frac{\partial q}{\partial\mathbf{v}_{2}}-\frac{\partial p}{\partial\mathbf{v}_{1}}\frac{\partial q}{\partial\mathbf{v}_{2}}-\frac{\partial p}{\partial\mathbf{v}_{2}}\frac{\partial q}{\partial\mathbf{v}_{1}}\right\} f(\mathbf{v}_{1})f(\mathbf{v}_{2}),\label{eq:Q}
\end{multline}
with
\begin{equation}
\mathbf{A}\left[f\right]\left(\mathbf{k},\omega,\mathbf{v}_{1},\mathbf{v}_{2}\right)=\pi\frac{\mathbf{k}\mathbf{k}\hat{W}(\mathbf{k})^{2}}{\left|\varepsilon\left[f\right]\left(\mathbf{k},\omega\right)\right|^{2}}\delta\left(\omega-\mathbf{k}.\mathbf{v}_{1}\right)\delta\left(\omega-\mathbf{k}.\mathbf{v}_{2}\right).\label{eq:tensor_A}
\end{equation}
We note that the tensor $\mathbf{A}$ is related to the tensor $\mathbf{B}$
of the Balescu--Guernsey--Lenard equation (\ref{eq:BLG_eq}):
\[
\mathbf{B}\left[f\right]\left(\mathbf{v}_{1},\mathbf{v}_{2}\right)=\frac{1}{L^{3}}\sum_{\mathbf{k}}\int\text{d}\omega\,\mathbf{A}\left[f\right]\left(\mathbf{k},\omega,\mathbf{v}_{1},\mathbf{v}_{2}\right),
\]
and that it shares all of its properties: it is symmetric as a tensor,
it is symmetric in its velocities argument 
\[\mathbf{A}\left(\mathbf{k},\omega,\mathbf{v}_{1},\mathbf{v}_{2}\right)=\mathbf{A}\left(\mathbf{k},\omega,\mathbf{v}_{1},\mathbf{v}_{2}\right)
\]
(momentum conservation), and we have 
\[\mathbf{A}\left(\mathbf{k},\omega,\mathbf{v}_{1},\mathbf{v}_{2}\right).\left(\mathbf{v}_{1}-\mathbf{v}_{2}\right)=0
\]
(energy conservation). These properties are related to the conservation
laws of the physical system, as we will see in section \ref{subsec:Conservation-laws}.
Using $\varepsilon\left[f\right]\left(-\mathbf{k},-\omega\right)=\varepsilon^{*}\left[f\right]\left(\mathbf{k},\omega\right)$,
we also have 
\[\mathbf{A}\left[f\right]\left(\mathbf{k},\omega,\mathbf{v}_{1},\mathbf{v}_{2}\right)=\mathbf{A}\left[f\right]\left(-\mathbf{k},-\omega,\mathbf{v}_{1},\mathbf{v}_{2}\right).
\]
$\mathbf{A}$ also has a symmetry property related to the time reversal
symmetry. Recalling that $I\left[f\right](\mathbf{v})=f\left(-\mathbf{v}\right)$
is the velocity inversion involution, we recall that $\varepsilon\left[I\left[f\right]\right]\left(\mathbf{k},-\omega\right)=\varepsilon^{*}\left[f\right]\left(\mathbf{k},\omega\right)$
and as a consequence 
\[ 
\mathbf{A}\left[I\left[f\right]\right]\left(\mathbf{k},-\omega,-\mathbf{v}_{1},-\mathbf{v}_{2}\right)=\mathbf{A}\left[f\right]\left(\mathbf{k},\omega,\mathbf{v}_{1},\mathbf{v}_{2}\right).
\]
We will discuss more deeply this property in section \ref{subsec:Time-reversal-symmetry,-quasi-po}.\\

Using equations (\ref{eq:H-decomp-1}), (\ref{eq:H_det}) and (\ref{eq:det_u})
we obtain an explicit formula for the large deviation Hamiltonian
\begin{equation}
H\left[f,p\right]=-\frac{1}{4\pi L^{3}}\sum_{\mathbf{k}}\int\text{d}\omega\,\log\left\{ 1-\mathcal{J}\left[f,p\right]\left(\mathbf{k},\omega\right)\right\} ,\label{eq:final_H}
\end{equation}
where $\mathcal{J}\left[f,p\right]\left(\mathbf{k},\omega\right)$
is defined in equations ((\ref{eq:J})-\ref{eq:Q}).

As a conclusion, in this section, we have established the path large
deviation principle
\[
\label{LDP}
\mathbf{P}\left(\left\{ f_{N}(\tau)\right\} _{0\leq\tau\leq T}=\left\{ f(\tau)\right\} _{0\leq\tau\leq T}\right)\underset{N\rightarrow\infty}{\asymp}\text{e}^{-NL^{3}\int_{0}^{T}\text{d}\tau\,\text{Sup}_{p}\left\{ \int\text{d}\mathbf{v}\,\dot{f}p-H[f,p]\right\} }\text{e}^{-NI_{0}\left[f\left(\tau=0\right)\right]},
\]
where $H$ is given by (\ref{eq:final_H}) and where $\tau=t/N$. \\

{\bf Density-current formulation of the large deviation principle.} We define the current as 
\[
\mathbf{j}_{N}\left(\mathbf{v},t\right)=-\frac{1}{NL^{3}}\int\text{d}\mathbf{r}\,\left(\frac{\partial V\left[\delta g_{N}\right]}{\partial\mathbf{r}}\delta g_{N}\right).
\]
In appendix \ref{sec:Current-formulation}, we prove that the large
deviation principle (\ref{LDP}) is equivalent to a empirical density-current formulation:
\[
\label{densitycurrent}
\mathbf{P}\left(\left\{ f_{N}(\tau),\mathbf{j}_{N}\left(\tau\right)\right\} _{0\leq t\leq T}=\left\{ f(\tau),\mathbf{j}\left(\tau\right)\right\} _{0\leq t\leq T}\right)\underset{N\rightarrow\infty}{\asymp}\text{e}^{-N\mathcal{A}\left[f,\mathbf{j}\right]}\text{e}^{-NI_{0}\left[f\left(\tau=0\right)\right]},
\]
where $\mathbf{j}_{N}\left(\tau\right)$
should be interpreted as
a time-averaged current after time rescaling, with
\[
\mathcal{A}\left[f,\mathbf{j}\right]=\begin{cases}
\begin{array}{ll}
L^{3}\int_{0}^{T}\text{d}\tau\,\tilde{L}\left[f,\mathbf{j}\right] & \text{if\,\,\,}\text{\ensuremath{\dot{f}}+\ensuremath{\frac{\partial}{\partial\mathbf{v}}\cdot}\ensuremath{\mathbf{j}}=0,}\\
+\infty & \text{otherwise}.
\end{array}\end{cases}
\]
and where $\tilde{L}\left[f,\mathbf{j}\right]=\underset{\mathbf{E}}{\text{Sup}}\left\{ \int\text{d}\mathbf{v}\,\mathbf{j}\cdot\mathbf{E}-\tilde{H}[f,\mathbf{E}]\right\} ,$
and $\tilde{H}$ is defined by $H\left[f,p\right]=\tilde{H}\left[f,\partial p/\partial\mathbf{v}\right]$. 

\section{Properties of the large deviation Hamiltonian\label{sec:Properties-of-the}}

In this section we check that the large deviation Hamiltonian (\ref{eq:final_H})
satisfies all the expected symmetry properties. In section \ref{subsec:Conservation-laws},
we check that the Hamiltonian (\ref{eq:final_H}) is consistent with
the mass, momentum and energy conservation laws. In section \ref{subsec:Time-reversal-symmetry,-quasi-po},
we show that the Hamiltonian (\ref{eq:final_H}) has a time-reversal
symmetry, and has the negative of the entropy, with conservation law
constraints and up to constants, as a quasipotential.

\subsection{Conservation laws \label{subsec:Conservation-laws}}

It is a classical exercise to prove that any conservation law is equivalent
to a symmetry property of the large deviation Hamiltonian, see for
instance section 7.3.2 of \cite{bouchet2020boltzmann}. From section
7.3.2 of \cite{bouchet2020boltzmann}, we know that a functional $C[f]$
is a conserved quantity of the large deviation principle (\ref{eq:PGD_H_GE})
if and only if for any $f$ and $p$
\begin{equation}
\int\text{d}\mathbf{v}\,\frac{\delta H}{\delta p\left(\mathbf{v}\right)}\frac{\delta C}{\delta f\left(\mathbf{v}\right)}=0,\label{eq:casimir_loc}
\end{equation}
 or equivalently, if for any $f$, $p$ and $\alpha\in\mathbb{R}$:
\begin{equation}
H[f,p]=H\left[f,p+\alpha\frac{\delta C}{\delta f}\right].\label{eq:casimir_glob}
\end{equation}
We will need the expression of the functional derivative of the Hamiltonian
$H$ with respect to its conjugate momentum $p$ throughout this section.
It reads
\begin{equation}
\frac{\delta H}{\delta p\left(\mathbf{v}\right)}[f,p]=\frac{1}{4\pi L^{3}}\sum_{\mathbf{k}}\int\text{d}\omega\,\frac{\frac{\delta\mathcal{J}}{\delta p\left(\mathbf{v}\right)}[f,p]\left(\mathbf{k},\omega\right)}{1-\mathcal{J}\left[f,p\right]\left(\mathbf{k},\omega\right)},\label{eq:functional_der_H}
\end{equation}
with
\small
\begin{equation}
\frac{\delta\mathcal{J}}{\delta p\left(\mathbf{v}\right)}[f,p]\left(\mathbf{k},\omega\right)=-4\pi\int\text{d}\mathbf{v}_{2}\,\frac{\partial}{\partial\mathbf{v}}\left\{ \mathbf{A}\left(\mathbf{k},\omega,\mathbf{v},\mathbf{v}_{2}\right)\left[\frac{\partial f}{\partial\mathbf{v}_{2}}f(\mathbf{v})-\frac{\partial f}{\partial\mathbf{v}}f(\mathbf{v}_{2})+2f(\mathbf{v})f(\mathbf{v}_{2})\left(\frac{\partial p}{\partial\mathbf{v}}-\frac{\partial p}{\partial\mathbf{v}_{2}}\right)\right]\right\} .\label{eq:functional_der_I}
\end{equation}
\normalsize

\paragraph{Mass conservation.}

The conservation of the total mass $M\left[f\right]=\int\text{d}\mathbf{v}\,f$
is immediately visible from equation (\ref{eq:casimir_glob}) as $H$
only depends on the derivative of the conjugated momentum $p$.

\paragraph{Momentum conservation.}

We define the total momentum $\mathbf{P}\left[f\right]=\int\text{d}\mathbf{v}\,\mathbf{v}f.$
It follows that $\frac{\delta\mathbf{P}}{\delta f\left(\mathbf{v}\right)}=\mathbf{v}$.
Using equation (\ref{eq:functional_der_I}) and partial integration,
the relation
\[
\int\text{d}\mathbf{v}_{1}\frac{\delta\mathcal{J}}{\delta p\left(\mathbf{v}_{1}\right)}[f,p]\left(\mathbf{k},\omega\right)\frac{\delta\mathbf{P}}{\delta f\left(\mathbf{v}_{1}\right)}=0
\]
is a direct consequence of the symmetry $\mathbf{A}\left(\mathbf{k},\omega,\mathbf{v}_{1},\mathbf{v}_{2}\right)=\mathbf{A}\left(\mathbf{k},\omega,\mathbf{v}_{2},\mathbf{v}_{1}\right).$
Then, using the relation (\ref{eq:functional_der_H}) between the
functional derivatives of $\mathcal{J}$ and $H$, we obtain
\[
\int\text{d}\mathbf{v}_{1}\frac{\delta H}{\delta p\left(\mathbf{v}_{1}\right)}[f,p]\left(\mathbf{k},\omega\right)\frac{\delta\mathbf{P}}{\delta f\left(\mathbf{v}_{1}\right)}=0.
\]
We have thus checked that the large deviation principle conserves
momentum. 

The conservation of mass and momentum should have been expected as
momentum and mass conservations were already granted from the expression
of the Hamiltonian (\ref{eq:H_Gartner_ellis}), as a direct consequence
of mass and momentum conservations for $f_{N}$ that can be deduced
from either equation (\ref{eq:slow}) or equation \ref{eq:slow-1})
.

\paragraph{Energy conservation.}

We define the total kinetic energy $E\left[f\right]=\int\text{d}\mathbf{v}\,\frac{\mathbf{v}^{2}}{2}f.$
It follows that $\frac{\delta E}{\delta f\left(\mathbf{v}\right)}=\mathbf{v}^{2}/2$.
Using equation (\ref{eq:functional_der_I}) and partial integration,
one can check that the relation
\[
\int\text{d}\mathbf{v}_{1}\frac{\delta\mathcal{J}}{\delta p\left(\mathbf{v}_{1}\right)}[f,p]\left(\mathbf{k},\omega\right)\frac{\delta E}{\delta f\left(\mathbf{v}_{1}\right)}=0
\]
is a direct consequence of the following symmetries of the tensor:
\[
\mathbf{A}\left(\mathbf{k},\omega,\mathbf{v}_{1},\mathbf{v}_{2}\right).\left(\mathbf{v}_{1}-\mathbf{v}_{2}\right)=0,
\]
and 
\[\mathbf{A}\left(\mathbf{k},\omega,\mathbf{v}_{1},\mathbf{v}_{2}\right)=\mathbf{A}\left(\mathbf{k},\omega,\mathbf{v}_{2},\mathbf{v}_{1}\right).
\] 
Then, using the relation (\ref{eq:functional_der_H}) between the
functional derivatives of $\mathcal{J}$ and $H$, we obtain
\[
\int\text{d}\mathbf{v}_{1}\frac{\delta H}{\delta p\left(\mathbf{v}_{1}\right)}\left(\mathbf{k},\omega\right)\frac{\delta E}{\delta f\left(\mathbf{v}_{1}\right)}=0.
\]
From the result (\ref{eq:casimir_loc}) we deduce that the large deviation
principle conserves the kinetic energy. 

The conservation of the kinetic energy is not a trivial consequence
of equation (\ref{eq:slow}) or equation (\ref{eq:slow-1}). Indeed,
from equation (\ref{eq:slow}) or equation (\ref{eq:slow-1}), at
any time some energy can be exchanged between the kinetic part $\int\text{d}\mathbf{v}\,\frac{\mathbf{v}^{2}}{2}f_{N}$
and the potential part related to $\delta g_{N}$. However $\int_{0}^{T}\text{d}t\int\text{d}\mathbf{v}\,\frac{\mathbf{v}^{2}}{2}\frac{\partial f_{N}}{\partial t}$
is equal to the negative of the variations of the potential energy.
Then, over any time $T$, these variations should remain bounded,
for the system to stay close to the set of homogenous solutions. As
a consequence, in accordance with our hypothesis of spatial homogeneity,
$\underset{T\rightarrow\infty}{\lim}\frac{1}{T}\int_{0}^{T}\text{d}t\int\text{d}\mathbf{v}\,\frac{\mathbf{v}^{2}}{2}\frac{\partial f_{N}}{\partial t}=0$.
This is the reason why we should have expected the conservation of
kinetic energy by the large deviation principle. The conservation
of kinetic energy by the large deviation principle, which is a conservation
for the slow effective dynamics for the empirical density, should
thus be interpreted as a conservation for time averages for the fast
process. If the system became inhomogeneous, this conservation
could be broken.

\subsection{Time-reversal symmetry, quasipotential, and entropy\label{subsec:Time-reversal-symmetry,-quasi-po}}

For the Hamiltonian dynamics (\ref{eq:dynamics}), we consider the
microcanonical measure with fixed energy $E$ and momentum fixed and
equal to zero, and denote $\mathbb{E}_{m}$ averages with respect
to the microcanonical measure. We expect the stationary probability
to observe $f_{N}=f$, to satisfy a large deviation principle 

\begin{equation}
\mathbb{E}_{m}\left[\delta\left(f_{N}-f\right)\right]\underset{N\rightarrow\infty}{\asymp}\exp\left\{ -NU\left[f\right]\right\} ,\label{eq:Quasipotential_Definition}
\end{equation}
where this large deviation principle defines the quasipotential $U$.

From classical equilibrium statistical mechanics considerations, for
this system with long range interactions, it is easy to justify that
the quasipotential is 
\begin{equation}
U\left[f\right]=\begin{cases}
\begin{array}{ll}
-\frac{S\left[f\right]}{k_{B}}+\frac{S_{m}(E)}{k_{B}} & \text{if\,\,\,}\int\text{d}\ensuremath{\mathbf{v}}\,f=1,\,\,\,\int\text{d}\ensuremath{\mathbf{v}}\,\mathbf{v}f=0,\,\,\,\text{and}\,\,\,\int\text{d}\ensuremath{\mathbf{v}}\,\frac{\ensuremath{\mathbf{v}^{2}}}{2}=E;\\
+\infty & \text{otherwise},
\end{array}\end{cases}\label{eq:Quasipotential}
\end{equation}
where

\[
S\left[f\right]=-k_{B}\int\text{d}\mathbf{v}\,f\log f
\]
is the entropy of the macrostate $f$ and
\[
S_{m}(E)=-k_{B}\inf_{f}\left\{ \int f\log f\left|\int\text{d}\ensuremath{\mathbf{v}}\,f=1,\,\,\,\int\text{d}\ensuremath{\mathbf{v}}\,\mathbf{v}f=0,\,\,\,\text{and}\,\,\,\int\text{d}\ensuremath{\mathbf{v}}\,\frac{\ensuremath{\mathbf{v}^{2}}}{2}=E\right.\right\} .
\]
is the equilibrium entropy. We have $S_{m}(E)=k_{B}\left[3\log(E)/2+3\log(4\pi)/2+3/2\right]$. 

It is also classically known that the Hamiltonian dynamics (\ref{eq:dynamics})
is time-reversible: the dynamics is symmetric by the change of variable
$(t,\mathbf{r}_{n},\mathbf{v}_{n})\rightarrow(-t,\mathbf{r}_{n},-\mathbf{v}_{n})$.
This is equivalent to say that if $\left\{ \mathbf{r}_{n}(t),\mathbf{v}_{n}(t)\right\} _{t\in\left[0,T\right]}$
is a solution of the Hamiltonian dynamics, then $\left\{ \mathbf{r}_{n}(T-t),-\mathbf{v}_{n}(T-t)\right\} _{t\in\left[0,T\right]}$
is also a solution. In order to take into account the change of sign
for the velocity, we define the linear operator on the set of function
of the velocity $I\left[f\right]\left(\mathbf{v}\right)=f\left(-\mathbf{v}\right)$.
We note that $I$ is an involution: $I^{2}=\text{Id}$. From the time
reversal symmetry for the Hamiltonian dynamical system, it is straightforward
to conclude that the stochastic process for the empirical density
$f_{N}$ should verify a generalized detailed balance symmetry. This
symmetry writes 
\small
\begin{equation}
\mathbf{P}_{T}\left(f_{N}(T)=f_{2}\left|f_{N}(0)=f_{1}\right.\right)\mathbf{P}_{m}\left(f_{N}=f_{1}\right)=\mathbf{P}_{T}\left(f_{N}(T)=I\left[f_{2}\right]\left|f_{N}(0)=I\left[f_{2}\right]\right.\right)\mathbf{P}_{m}\left(f_{N}=I\left[f_{2}\right]\right),\label{eq:detailed balance empirical density}
\end{equation}
\normalsize
where $\mathbf{P}_{m}$ is the stationary measure with respect to
the microcanonical measure, $\mathbf{P}_{T}$ are the transition probabilities
for the microcanonical measure. The term ``generalized'' means that
the symmetry holds using the involution $I$. It is a classical exercise,
see for instance section 7.3.2 of \cite{bouchet2020boltzmann}, to
prove that the detailed balance condition (\ref{eq:detailed balance empirical density})
implies a detailed balance symmetry at the level of the Hamiltonian:
for any $f$ and $p$, 

\begin{equation}
H\left[I\left[f\right],-I\left[p\right]\right]=H\left[f,p+\frac{\delta U}{\delta f}\right].\label{eq:Generalized_Detailed_Balance_Hamiltonian}
\end{equation}
From the relation (\ref{eq:Quasipotential}) between the quasipotential
$U$ and the entropy $S$, using the conservation law symmetries of
the large deviation Hamiltonian (\ref{eq:casimir_glob}) we can conclude
that the generalized detailed balance symmetry (\ref{eq:Generalized_Detailed_Balance_Hamiltonian})
is equivalent to the symmetry: for any $f$ and $p$,

\begin{equation}
H\left[I\left[f\right],-I\left[p\right]\right]=H\left[f,p-\frac{1}{k_{B}}\frac{\delta S}{\delta f}\right].\label{eq:Generalized_Detailed_Balance_Hamiltonian-1}
\end{equation}
One may directly check this symmetry, from (\ref{eq:final_H}), using
the time reversal symmetry for $A$: 
\[
\mathbf{A}\left[I\left[f\right]\right]\left(\mathbf{k},-\omega,-\mathbf{v}_{1},-\mathbf{v}_{2}\right)=\mathbf{A}\left[f\right]\left(\mathbf{k},\omega,\mathbf{v}_{1},\mathbf{v}_{2}\right).
\]
It is however simpler to first note that for spatially homogeneous
systems, which is the case in this paper, one has the further symmetry
: 
\[
H\left[I\left[f\right],I\left[p\right]\right]=H\left[f,p\right].
\]
This symmetry can be checked starting from (\ref{eq:final_H}) and
(\ref{eq:J}), using 
\[\mathbf{A}\left[I\left[f\right]\right]\left(\mathbf{k},-\omega,-\mathbf{v}_{1},-\mathbf{v}_{2}\right)=\mathbf{A}\left[f\right]\left(\mathbf{k},\omega,\mathbf{v}_{1},\mathbf{v}_{2}\right),
\]
to conclude that 
\[\mathcal{J}\left[I\left[f\right],I\left[p\right]\right]\left(\mathbf{k},\omega\right)=\mathcal{J}\left[f,p\right]\left(\mathbf{k},-\omega\right).
\]
With this remark, we can conclude that the generalized detailed balance
condition is equivalent to: for any $f$ and $p$, 
\begin{equation}
H\left[f,-p\right]=H\left[f,p-\frac{1}{k_{B}}\frac{\delta S}{\delta f}\right].\label{eq:Detailed_Balance_Hamiltonian}
\end{equation}
This last condition is a detailed balance condition at the level of
large deviations (see for instance section 7.3.2 of \cite{bouchet2020boltzmann}).
In order to check directly (\ref{eq:Detailed_Balance_Hamiltonian}),
one can start from (\ref{eq:final_H}) and (\ref{eq:J}), and see
that this follows from $\mathcal{J}\left[f,p-k_{B}^{-1}\delta S/\delta f\right]-\mathcal{J}\left[f,-p\right]=0$.
One can see that this last equality is equivalent to the relation:
for any $f$ and $p$, $\mathcal{L}\left[f,p\right]=Q\left[f,p,k_{B}^{-1}\delta S/\delta f\right],$
using (\ref{eq:J}) and that $\mathcal{L}$ is linear and $\mathcal{Q}$
quadratic with respect to $p$. Using (\ref{eq:L}) and (\ref{eq:Q})
and $\partial/\partial\mathbf{v}(\delta S/\delta f)/k_{B}=1/f\partial f/\partial\mathbf{v}$,
this is easily verified using $\text{\ensuremath{\mathbf{A}}}\left[f\right]\left(\mathbf{k},\omega,\mathbf{v}_{1},\mathbf{v}_{2}\right)=\text{\ensuremath{\mathbf{A}\left[f\right]\left(\mathbf{k},\omega,\mathbf{v}_{2},\mathbf{v}_{1}\right)}}$.

As a final remark, we note that the quasipotential and the entropy
are solutions to the stationary Hamilton-Jacobi equation
\[
H\left[f,\frac{\delta U}{\delta f}\right]=H\left[f,-\frac{1}{k_{B}}\frac{\delta S}{\delta f}\right]=0.
\]
Those are direct consequences of any of the detailed balance symmetries:
(\ref{eq:Detailed_Balance_Hamiltonian}), (\ref{eq:Generalized_Detailed_Balance_Hamiltonian})
or (\ref{eq:Generalized_Detailed_Balance_Hamiltonian-1}). \\

In this section we have explained that $U$ (\ref{eq:Quasipotential})
is the quasipotential. We have argued that the large deviation Hamiltonian
satisfies the generalized detailed balance symmetry (\ref{eq:Generalized_Detailed_Balance_Hamiltonian-1})
as a consequence of the microscopic time reversibility, and checked
directly this relation from the explicit Hamiltonian equations. We
have moreover justified that that the large deviation Hamiltonian
satisfies the detailed balance symmetry (\ref{eq:Detailed_Balance_Hamiltonian}).
This proves that $U$ satisfies the stationary Hamilton-Jacobi equation.

\subsection{A remark on Gaussian approximations of the large deviation principle
and cumulant expansions\label{subsec:Gaussian-Approximation}}

In a recent paper \cite{2021arXiv210104455F} dedicated to the path
large deviations for homogeneous plasma, for fluctuations on scales
$k\lambda_{D}\gg1$ where $\lambda_{D}$ is the Debye length and $k$
a wavenumber, we obtained a large deviation Hamiltonian which is quadratic
in $p$ featuring locally Gaussian distribution of the large deviations.
We have obtained this quadratic in $p$ Hamiltonian, either from
the large deviation Hamiltonian associated with the Boltzmann equation,
or from an expansion using $k\lambda_{D}\gg1$ of the expression (\ref{eq:H_Gartner_ellis}).
The technical approach in \cite{2021arXiv210104455F} is different. In~ \cite{2021arXiv210104455F} we used a tedious cumulant expansion. This paper   rather uses the Szegö--Widom theorem, a very efficient approach.

The present paper considers any interaction, and not just the Coulomb
interaction case, but it is clear that the result in \cite{2021arXiv210104455F}
should be recovered from the results of the present paper. In appendix
\ref{sec:Consistence-with-the}, we show that the large deviation
Hamiltonian of this paper (\ref{eq:final_H}), is fully consistent
with our previous perturbative result for $k\lambda_{D}\gg1$. We
also justify an assumption we used in \cite{2021arXiv210104455F}
about the cumulant series expansion of (\ref{eq:H_Gartner_ellis})
from the formula (\ref{eq:final_H}) for the large deviation Hamiltonian
associated with the Balescu--Guernsey--Lenard equation.

\section{Perspectives\label{sec:Perspectives}}

The main result of this paper is the derivation of a large deviation
principle (\ref{eq:Large_Deviation_Principle_Intro}), for the velocity
empirical density, for the Hamiltonian dynamics of $N$ particles
which interact through mean-field interactions. We have
obtained an explicit formula for the large deviation Hamiltonian (\ref{eq:LDP_Hamiltonien_Intro}-\ref{eq:LDP_J_Intro})
and we have checked all its symmetry properties. This result opens
many mathematical and theoretical questions, as well as interesting
applications. 

This large deviation result relies on natural assumptions. Some of
these assumptions are also required to establish the Balescu--Guernsey--Lenard
kinetic equation, but the hypotheses made to obtain the large deviation
principle seem stronger. The first assumption is the validity of the
quasilinear approximation: we neglected non linear terms of order
$1/\sqrt{N}$ in the equation for the fluctuations of the empirical
density. This amounts to neglecting
possible effects of large deviations of the fluctuations, and describing
the fluctuation process at a Gaussian level only. The second assumption is the convergence of the process of
fluctuations to a stationary Gaussian process and more specifically
the convergence of the large time asymptotics for the large deviation
estimates over this process. A proof would also require the study of the mixing properties for
this Gaussian processes. The mixing properties are critical to justify
the Markov behavior described by the slow-fast large deviation principle.
While the proofs of these assumptions are beyond the scope of this
paper, they open very interesting questions for both theoretical physicists
and mathematicians. 

Systems with long range interactions are important for many phenomena.
However, more elaborate models than the one we used in this paper
could be more appropriate to describe physical situations where rare
event are important for applications. Of special interest,
would be the derivation of large deviation principles for inhomogeneous
systems with long range interactions, for instance self-gravitating
systems. This is an exciting application, that would open the way
to the study of the rare destabilization of globular clusters or galaxies, or the formation of inhomogeneous structures of smaller scales in self gravitating systems.
Another very interesting generalization would be for the dynamics
of $N$ point-vortices for two-dimensional hydrodynamics. Another
generalization should also consider dynamics of particles driven by
stochastic forces, which generically lead to irreversible stochastic
processes. For those systems, explicit results for the large deviation
theory would be extremely useful for explaining non-equilibrium phase
transitions in two dimensional \cite{Bouchet_Simonnet_2008} and geostrophic
turbulence \cite{BOUCHET:2019:C}, or in systems with long range interactions
\cite{Nardini_Gupta_Ruffo_Dauxois_Bouchet_2012_kinetic,NardiniGuptaBouchet-2012-JSMTE}.

\appendix

\section{Long time large deviations for quadratic observables of Gaussian
processes, functional determinants and the Szegö--Widom theorem for
Fredholm determinants\label{sec:Long-time-large}}

In this appendix, we explain how we can use the Szegö--Widom theorem
in order to evaluate the large time asymptotics of Fredholm determinants
that appears when computing the cumulant generating function of a
quadratic observable of a Gaussian process. We follow the ideas in \cite{Bouchet-Roger-Zaboronski}, adapting the discussion for the case of Gaussian processes with complex variables. 

Let $Y_{t}$ be a stationary $\mathbb{C}^{n}$-valued Gaussian
process with correlation matrix $C\left(t\right)=\mathbb{E}\left(Y_{t}\otimes Y_{0}^{*}\right)$ and with a zero relation matrix $R(t)=\mathbb{E}\left(Y_{t}\otimes Y_{0}\right)=0$, let
$M\in\mathcal{M}_{n}\left(\mathbb{C}\right)$ be a $n\times n$ Hermitian
matrix. The aim of this appendix is to prove that 
\begin{equation}
\log\mathbb{E}\exp\left(\int_{0}^{T}\text{d}t\,Y_{t}^{*\intercal}MY_{t}\right)\underset{T\rightarrow\infty}{\sim}-\frac{T}{2\pi}\int\text{d}\omega\,\log\det\left(I_{n}-M\tilde{C}\left(\omega\right)\right),\label{eq:asymptotics_quadratic_observable}
\end{equation}
where $\tilde{C}\left(\omega\right)=\int_{\mathbb{R}}\text{e}^{i\omega t}C\left(t\right)\text{d}t$
is the Fourier transform of the correlation matrix $C\left(t\right)$
and $I_{n}$ is the $n\times n$ identity matrix. We note that the
determinant of the r.h.s. of (\ref{eq:asymptotics_quadratic_observable})
is a real number. Indeed, as $Y_{t}$ is a stationary process, $\tilde{C}\left(\omega\right)$
and $M$ are Hermitian matrices, then the determinant is the determinant
of a Hermitian operator and is a real number.

For pedagogical reasons, in this appendix the result (\ref{eq:asymptotics_quadratic_observable})
is stated for a process $Y_{t}$ that takes values in a finite-dimensional
space. However with adapted hypotheses, this result can be generalized
when $Y_{t}$ is a stationary $\mathcal{H}$-valued Gaussian process,
where $\mathcal{H}$ is a Hilbert space, and where $M$ is a Hermitian
operator on $\mathcal{H}$. 

In section \ref{subsec:The-Szeg=0000F6=002013Widom-theorem}, we state
the Szegö--Widom theorem. In section \ref{subsec:Expectation-of-functionals},
we explain that the left hand side of (\ref{eq:asymptotics_quadratic_observable})
is the log of the determinant of a Gaussian integral, that this quantity
can be expressed as a functional determinant for linear operators
on $L^{2}\left(\left[0,T\right],\mathbb{C}^{n}\right)$, and that
Szegö--Widom theorem reduces it to the computation of frequency integrals
of determinants of operators on the space $\mathbb{C}^{n}$, as expressed
by (\ref{eq:asymptotics_quadratic_observable}).

\subsection{The Szegö--Widom theorem \label{subsec:The-Szeg=0000F6=002013Widom-theorem}}

We first define integral operators on $L^{2}\left(\left[0,T\right],\mathbb{C}^{n}\right)$.
We considers maps $\varphi:\left[0,T\right]\rightarrow\mathbb{C}^{n}$
and $K:\mathbb{R}\rightarrow\mathcal{M}_{n}\left(\mathbb{C}\right)$,
where $\mathcal{M}_{n}\left(\mathbb{C}\right)$ is the set of $n\times n$
complex matrices. We define the integral operator $\mathbf{K}_T$ by
\begin{equation}
\mathbf{K}_T\varphi\left(t\right)=\int_{0}^{T}K\left(t-s\right)\varphi\left(s\right)\text{d}s,\label{eq:Integral_Operator}
\end{equation}
$\mathbf{K}_T$ is a linear operator of $L^{2}\left(\left[0,T\right],\mathbb{C}^{n}\right)$.
$K$ is called the kernel of the operator $\mathbf{K}_T$. 

The Szegö--Widom theorem allows to compute large $T$ asymptotics
of the logarithm of the Fredholm determinant of the integral operator
$\text{Id}+\mathbf{K}_T$. The result is
\begin{equation}
\log\det_{\left[0,T\right]}\left(\text{Id}+\mathbf{K}_T\right)\underset{T\rightarrow\infty}{\sim}\frac{T}{2\pi}\int\text{d}\omega\,\log\det\left(I_{n}+\int_{\mathbb{R}}\text{e}^{i\omega t}K\left(t\right)\text{d}t\right),\label{eq:Szego-Widom-th}
\end{equation}
where $I_{n}$ is the $n\times n$ identity matrix. Whereas the determinant
on the l.h.s. of this expression, denoted by the subscript $\left[0,T\right]$
is a Fredholm determinant, the determinant on the r.h.s. is a matrix
determinant which can be more easily computed. Further details about
this theorem and its possible applications can be found in \cite{Bouchet-Roger-Zaboronski}.

\subsection{Expectation of functionals of Gaussian processes\label{subsec:Expectation-of-functionals}}

Let $Y_{t}$ be a $\mathbb{C}^{n}$-valued stationary Gaussian process
with correlation matrix
\[
C\left(t\right)=\mathbb{E}\left(Y_{t}\otimes Y_{0}^{*}\right),
\]
and with zero relation matrix
\[
R(t)=\mathbb{E}\left(Y_{t}\otimes Y_{0}\right)=0.
\]
We will compute the large time asymptotics of
\[
\mathcal{U}\left(T\right)=\log\mathbb{E}\exp\left(\int_{0}^{T}\text{d}t\,Y_{t}^{*\intercal}\mathbf{M}_T Y_{t}\right),
\]
where $\mathbf{M}_T$ is an integral operator on $L^{2}\left(\left[0,T\right],\mathbb{C}^{n}\right)$
whose integral kernel is given by $M\left(t\right)$ (see the definition
\eqref{eq:Integral_Operator}). We assume that for all times $t$, $M\left(t\right)$
is a $n\times n$ Hermitian matrix. As $Y_{t}$ is a Gaussian process
we can compute the expectation as a Gaussian integral. It is straightforward
to check that
\[
\mathbb{E}\exp\left(\int_{0}^{T}\text{d}t\,Y_{t}^{*\intercal}\mathbf{M}_T Y_{t}\right)=\det_{\left[0,T\right]}\left(\text{Id}-(\mathbf{M} \mathbf{C})_T \right)^{-1},
\]
where $(\mathbf{MC})_T$ is the integral operator whose kernel is $(M\star C)(t)$ the convolution product on $[0,T]$ of the kernels $M(t)$ and $C(t)$.

Then, we can deduce the following expression for $u$
\[
\mathcal{U}\left(T\right)=-\log\det_{\left[0,T\right]}\left(\text{Id}-(\mathbf{M}\mathbf{C})_T \right),
\]
where the determinant is the Fredholm determinant of the integral
operator $\text{Id}-(\mathbf{M}\mathbf{C})_T$. Generally, it is not
obvious how to compute this kind of Fredholm determinant. Fortunately,
we can use the Szegö--Widom theorem to obtain an expression for large
$T$ asymptotics as a finite-dimensional determinant. Using the result
(\ref{eq:Szego-Widom-th}) from section \ref{subsec:The-Szeg=0000F6=002013Widom-theorem},
we get
\[
\mathcal{U}\left(T\right)\underset{T\rightarrow\infty}{\sim}-\frac{T}{2\pi}\int\text{d}\omega\,\log\det\left(I_{n}-\int_{\mathbb{R}}\text{e}^{-i\omega t}\left(M\star C\right)\left(t\right)\text{d}t\right).
\]
In the special case when $\mathbf{M}_T$ is a diagonal integral operator,
i.e. when its kernel is $M(t)=M\delta(t)$, we can write
\[
\mathcal{U}\left(T\right)\underset{T\rightarrow\infty}{\sim}-\frac{T}{2\pi}\int\text{d}\omega\,\log\det\left(I_{n}-M\int_{\mathbb{R}}\text{e}^{-i\omega t}C\left(t\right)\text{d}t\right),
\]
which is the result (\ref{eq:asymptotics_quadratic_observable}).
In these expressions, the determinant to be computed on the r.h.s.
is the determinant of a $n\times n$ matrix.

\section{Computation of the determinant of the operator $u_{\mathbf{k},\omega}$\label{sec:determinant}}

In this appendix, we compute the determinant of the operator $u_{\mathbf{k},\omega}$,
encountered in section \ref{subsec:Application-of-the}, and defined
by
\begin{eqnarray*}
u_{\mathbf{k},\omega}\left[\varphi\right]\left(\mathbf{v}_{1}\right) & = & \varphi\left(\mathbf{v}_{1}\right)-\int\text{d}\mathbf{v}_{2}\text{d}\mathbf{v}_{3}\,M\left(\mathbf{k},\mathbf{v}_{1},\mathbf{v}_{2}\right)\widetilde{\mathcal{C}_{GG}}\left(\mathbf{k},\omega,\mathbf{v}_{2},\mathbf{v}_{3}\right)\varphi\left(\mathbf{v}_{3}\right),
\end{eqnarray*}
for any $\varphi\in\mathcal{H}_{\mathbf{v}}$, $\mathcal{H}_{\mathbf{v}}$
being the Hilbert space of complex functions over the velocity space.
Using equation (\ref{eq:kernel-m}), we can simplify this expression
\begin{equation}
u_{\mathbf{k},\omega}\left[\varphi\right]\left(\mathbf{v}_{1}\right)=\varphi\left(\mathbf{v}_{1}\right)-i\hat{W}\left(\mathbf{k}\right)\mathbf{k}\cdot\int\text{d}\mathbf{v}_{2}\text{d}\mathbf{v}_{3}\,\widetilde{\mathcal{C}_{GG}}\left(\mathbf{k},\omega,\mathbf{v}_{2},\mathbf{v}_{3}\right)\left\{ \frac{\partial p}{\partial\mathbf{v}}\left(\mathbf{v}_{2}\right)-\frac{\partial p}{\partial\mathbf{v}}\left(\mathbf{v}_{1}\right)\right\} \varphi\left(\mathbf{v}_{3}\right).\label{eq:u_explicit}
\end{equation}

We note that the operator $u_{\mathbf{k},\omega}$ has the form 
\begin{equation}
u_{\mathbf{k},\omega}:\varphi\longmapsto\varphi-\left\langle w,\mathbf{Q}\varphi\right\rangle v-\left\langle v,\mathbf{Q}\varphi\right\rangle w,\label{eq:u_abstrait}
\end{equation}
where $\mathbf{Q}$ is a Hermitian operator over $\mathcal{H_{\mathbf{v}}}$,
$w$ and $v$ are complex functions over the velocity space, and $\left\langle .,.\right\rangle $
denotes the Hermitian product: $\left\langle a,b\right\rangle =\int\text{d}\mathbf{v}\,a^{*}\left(\mathbf{v}\right)b\left(\mathbf{v}\right)$.
The connection is made between formulas (\ref{eq:u_explicit}) and
(\ref{eq:u_abstrait}) by setting $v\left(\mathbf{v}\right)=-i\mathbf{k}.\frac{\partial p}{\partial\mathbf{v}}$
, $w\left(\mathbf{v}\right)=\hat{W}\left(\mathbf{k}\right)$ and $\mathbf{Q}\left[\phi\right]\left(\mathbf{v}_{1}\right)=\int\text{d}\mathbf{v}_{2}\,\widetilde{\mathcal{C}_{GG}}\left(\mathbf{k},\omega,\mathbf{v}_{1},\mathbf{v}_{2}\right)\phi\left(\mathbf{v}_{2}\right)$.
Using (\ref{eq:Symetrie_CGG}), we see that $Q$ is a Hermitian operator.
We note that, whenever $\frac{\partial p}{\partial\mathbf{v}}$ is
not a constant in the velocity space, $v$ and $w$ are linearly independent. 

Formula (\ref{eq:u_abstrait}) shows that $u_{\mathbf{k},\omega}-\text{Id}$
is a rank two linear operator. Then $\det_{\mathcal{H}_{\mathbf{v}}}u_{\mathbf{k},\omega}$
is the determinant of the operator $u_{\mathbf{k},\omega}$ restricted
to $\text{span\ensuremath{\left(v,w\right)}}$: 
\[
\det_{\mathcal{H}_{\mathbf{v}}}u_{\mathbf{k},\omega}=\begin{vmatrix}1-\left\langle w,\mathbf{Q}v\right\rangle  & -\left\langle w,\mathbf{Q}w\right\rangle \\
-\left\langle v,\mathbf{Q}v\right\rangle  & 1-\left\langle v,\mathbf{Q}w\right\rangle 
\end{vmatrix}.
\]
Then
\[
\det_{\mathcal{H}_{\mathbf{v}}}u_{\mathbf{k},\omega}=1-2\Re\left[\left\langle v,\mathbf{Q}w\right\rangle \right]+\left\langle v,\mathbf{Q}w\right\rangle \left\langle v,\mathbf{Q}w\right\rangle ^{*}-\left\langle w,\mathbf{Q}w\right\rangle \left\langle v,\mathbf{Q}v\right\rangle .
\]
where we have used $\left\langle w,\mathbf{Q}v\right\rangle =\left\langle \mathbf{Q}w,v\right\rangle =\left\langle v,\mathbf{Q}w\right\rangle ^{*}$,
as $\mathbf{Q}$ is an Hermitian operator. 

We can explicitly compute the determinant of (\ref{eq:u_explicit}).
We have
\[
\left\langle v,\mathbf{Q}v\right\rangle =\int\text{d}\mathbf{v}_{1}\text{d}\mathbf{v}_{2}\,\mathbf{k}.\frac{\partial p}{\partial\mathbf{v}_{1}}\mathbf{k}.\frac{\partial p}{\partial\mathbf{v}_{2}}\widetilde{\mathcal{C}_{GG}}\left(\mathbf{k},\omega,\mathbf{v}_{1},\mathbf{v}_{2}\right),
\]
\[
\left\langle v,\mathbf{Q}w\right\rangle =i\int\text{d}\mathbf{v}_{1}\,\mathbf{k}.\frac{\partial p}{\partial\mathbf{v}_{1}}\widetilde{\mathcal{C}_{VG}}\left(\mathbf{k},\omega,\mathbf{v}_{1}\right)^{*},
\]
and 
\[
\left\langle w,\mathbf{Q}w\right\rangle =\widetilde{\mathcal{C}_{VV}}\left(\mathbf{k},\omega\right),
\]
where $\widetilde{\mathcal{C}_{VG}}$, $\widetilde{\mathcal{C}_{VV}}$
and $\widetilde{\mathcal{C}_{GG}}$ are the two-point correlations
functions of the quasi-linear problem computed in section \ref{subsec:The-quasi-stationary-Gaussian},
and we have used (\ref{eq:relations_C_1}-\ref{eq:relations_C_2}).

We conclude that 
\begin{multline*}
\underset{\mathcal{H}_{\mathbf{v}}}{\det}\left(u_{\mathbf{k},\omega}\right)=1+2\int\text{d}\mathbf{v}_{1}\,\mathbf{k}.\frac{\partial p}{\partial\mathbf{v}_{1}}\Im\left(\widetilde{\mathcal{C}_{VG}}\left(\mathbf{k},\omega,\mathbf{v}_{1}\right)\right)\\
+\int\text{d}\mathbf{v}_{1}\text{d}\mathbf{v}_{2}\,\mathbf{k}.\frac{\partial p}{\partial\mathbf{v}_{1}}\mathbf{k}.\frac{\partial p}{\partial\mathbf{v}_{2}}\left\{ \widetilde{\mathcal{C}_{VG}}\left(\mathbf{k},\omega,\mathbf{v}_{1}\right)\widetilde{\mathcal{C}_{VG}}\left(\mathbf{k},\omega,\mathbf{v}_{2}\right)^{*}-\widetilde{\mathcal{C}_{VV}}\left(\mathbf{k},\omega\right)\widetilde{\mathcal{C}_{GG}}\left(\mathbf{k},\omega,\mathbf{v}_{1},\mathbf{v}_{2}\right)\right\} .
\end{multline*}

\section{Consistency and validation of the cumulant series expansion\label{sec:Consistence-with-the}}

In this appendix, we expand $H$ from the formula (\ref{eq:final_H})
in powers of $p$. This amounts at a cumulant expansion for the statistics
of the fluctuations. We use this expansion to prove a conjecture made
in the paper \cite{2021arXiv210104455F}, and to recover the Gaussian
(order two) truncation we computed in \cite{2021arXiv210104455F}.\\

We expand the logarithm in formula (\ref{eq:final_H}) to obtain 
\begin{equation}
H\left[f,p\right]=\frac{1}{4\pi L^{3}}\sum_{\mathbf{k}}\int\text{d}\omega\,\sum_{n=1}^{+\infty}\frac{1}{n}\left(\mathcal{J}\left[f,p\right]\left(\mathbf{k},\omega\right)\right)^{n}=\sum_{n=1}^{+\infty}H^{(n)}\left[f,p\right].\label{d=0000E9finition H^n}
\end{equation}
The second equality defines $H^{(n)}$ as being the terms homogeneous
of order $n$ in $p$ in this expansion. It is the $n$-th cumulant. 

We also define $\mathbf{B}^{(m)}$ as
\[
\mathbf{B}^{(m)}\left(\mathbf{v}_{1},\ldots,\mathbf{v}_{2m}\right)=\frac{\left(2\pi\right)^{2m}}{4\pi mL^{3}}\sum_{\mathbf{k}}\int_{\Gamma}\text{d}\omega\,\frac{W\left(\mathbf{k}\right)^{2m}}{\left|\varepsilon\left(\mathbf{k},\omega\right)\right|^{2m}}\mathbf{k}^{\otimes2m}\prod_{i=1}^{2m}\delta\left(\omega-\mathbf{k}.\mathbf{v}_{i}\right).
\]
$\mathbf{B}^{(m)}$ is a rank $2m$ tensor. $l^{(k)}$ and $q^{(k)}$
are defined by the relations. We have $\mathcal{J}\left[f,p\right]=\mathcal{L}\left[f,p\right]+Q\left[f,p,p\right],$
where $\mathcal{L}$ and $Q$ are defined in equations (\ref{eq:L})
and (\ref{eq:Q}). We will need to compute $\left(\mathcal{L}\left[f,p\right]\right)^{k}$,
which is $\mathcal{L}\left[f,p\right]$ to the power $k$. We define
$l^{(k)}$ and $q^{(k)}$ by 
\[
\left(\mathcal{L}\left[f,p\right]\right)^{k}=\int\text{d}\mathbf{v}_{1}\cdots\text{d}\mathbf{v}_{2k}\,l^{(k)}\left[f,p\right]\prod_{j=1}^{k}\mathbf{A}\left(\mathbf{k},\omega,\mathbf{v}_{2j-1},\mathbf{v}_{2j}\right),
\]
and
\[
\left(Q\left[f,p,p\right]\right)^{k}=\int\text{d}\mathbf{v}_{1}\cdots\text{d}\mathbf{v}_{2k}\,q^{(k)}\left[f,p,p\right]\prod_{j=1}^{k}\mathbf{A}\left(\mathbf{k},\omega,\mathbf{v}_{2j-1},\mathbf{v}_{2j}\right).
\]
$l^{(k)}$ and $q^{(k)}$ are both tensors of order $2k$. $l^{(k)}$
depends on $p$ as a homogeneous function of order $k$. $q^{(k)}$
depends on $p$ as a homogeneous function of order $2k$.

In the expansion of $\left(\mathcal{J}\left[f,p\right]\right)^{n}$
using $\mathcal{J}\left[f,p\right]=\mathcal{L}\left[f,p\right]+Q\left[f,p,p\right],$
we see that for all $m\in\left[n/2,n\right]\cap\mathbb{N}$, $\mathcal{L}^{2m-n}\left[f,p\right]Q^{n-m}\left[f,p,p\right]$
is homogeneous of order $n$ in $p$. Using this remark, from equation
(\ref{d=0000E9finition H^n}) we obtain 
\begin{equation}
H^{(n)}\left[f,p\right]=\sum_{m\in\left[n/2,n\right]\cap\mathbb{N}}\int\text{d}\mathbf{v}_{1}\cdots\text{d}\mathbf{v}_{2m}\,\left(\begin{array}{c}
m\\
2m-n
\end{array}\right)\frac{1}{\left(2\pi\right)^{2m}}\mathbf{B}^{(m)}\left(\mathbf{v}_{1},\ldots,\mathbf{v}_{2m}\right):l^{(2m-n)}\left[f,p\right]q^{(n-m)}\left[f,p,p\right],\label{eq:formula_H_n}
\end{equation}
where the symbol $":"$ means a contraction of a tensor of order $2m$
with another tensor of order $2m$. 

This result ensures that as soon as $n>2,$ $H^{(n)}$ only includes
terms proportional to the tensors $\mathbf{B}^{(m)}$ with $m\geq n/2\geq2.$
In \cite{2021arXiv210104455F}, we used a conjecture on $H^{(n)}$
to justify the Gaussian truncation of the cumulant series expansion
of (\ref{eq:H_Gartner_ellis}). The conjecture was that only the two
first cumulants $H^{(1)}$ and $H^{(2)}$ do involve the tensor $\mathbf{B}=\mathbf{B}^{(1)}$,
whereas all the other cumulants $H^{(n)}$ for $n>2$ only involve
tensors $\mathbf{B}^{(m)}$ with $m\geq2$. Expansion thus (\ref{eq:formula_H_n})
justifies this conjecture.

In \cite{2021arXiv210104455F}, from a cumulant series expansion,
we obtained that 
\[
H\left[f,p\right]=H_{\text{quad}}\left[f,p\right]+O\left(p^{2}\right),
\]
where
\begin{multline}
H_{\text{quad}}\left[f,p\right]=\int\text{d}\mathbf{v}_{1}\text{d}\mathbf{v}_{2}\,\mathbf{B}\left(\mathbf{v}_{1},\mathbf{v}_{2}\right):\frac{\partial p}{\partial\mathbf{v}_{1}}\left\{ \frac{\partial f}{\partial\mathbf{v}_{2}}f(\mathbf{v}_{1})-f(\mathbf{v}_{2})\frac{\partial f}{\partial\mathbf{v}_{1}}\right\} \\
+\int\text{d}\mathbf{v}_{1}\text{d}\mathbf{v}_{2}\,\mathbf{B}\left(\mathbf{v}_{1},\mathbf{v}_{2}\right):\left\{ \frac{\partial p}{\partial\mathbf{v}_{1}}\frac{\partial p}{\partial\mathbf{v}_{1}}-\frac{\partial p}{\partial\mathbf{v}_{1}}\frac{\partial p}{\partial\mathbf{v}_{2}}\right\} f(\mathbf{v}_{1})f(\mathbf{v}_{2})\\
+\int\text{d\ensuremath{\mathbf{v}_{1}}}\text{d}\mathbf{v}_{2}\text{d\ensuremath{\mathbf{v}_{3}}}\text{d}\mathbf{v}_{4}\,\mathbf{B}^{(2)}\left(\mathbf{v}_{1},\mathbf{v}_{2},\mathbf{v}_{3},\mathbf{v}_{4}\right):\frac{\partial p}{\partial\mathbf{v}_{1}}\frac{\partial p}{\partial\mathbf{v}_{2}}\left\{ f(\mathbf{v}_{1})f(\mathbf{v}_{2})\frac{\partial f}{\partial\mathbf{v}_{3}}\frac{\partial f}{\partial\mathbf{v}_{4}}\right. \\
\left. -2f(\mathbf{v}_{1})\frac{\partial f}{\partial\mathbf{v}_{2}}f(\mathbf{v}_{3})\frac{\partial f}{\partial\mathbf{v}_{4}}+\frac{\partial f}{\partial\mathbf{v}_{1}}\frac{\partial f}{\partial\mathbf{v}_{2}}f(\mathbf{v}_{3})f(\mathbf{v}_{4})\right\} ,\label{eq:quadratic_ham-1}
\end{multline}
and where $O\left(p^{2}\right)$ designates terms that are of order
more than two in the conjugate momentum $p$, and the symbol $":"$
means a contraction of two tensors of order 2 and 4, for the two first
lines and the third line, respectively. We see that $H^{(1)}+H^{(2)}=H_{\text{quad}}.$
We conclude that (\ref{eq:final_H}) is consistent with the quadratic
approximation of the large deviation Hamiltonian we obtained in \cite{2021arXiv210104455F}.

\section{Current formulation of the large deviation principle\label{sec:Current-formulation}}

Because the particle number is conserved, it is clear that the dynamics
of the empirical density has a conservative form $\frac{\partial f_{N}}{\partial t}+\frac{\partial}{\partial\mathbf{v}}\cdot\mathbf{j}_{N}=0$.
For the microscopic dynamics (before time averaging), this is a consequence
of equations (\ref{eq:slow}) or (\ref{eq:slow-1}) with\textbf{
\[
\mathbf{j}_{N}\left(\mathbf{v},t\right)=-\frac{1}{NL^{3}}\int\text{d}\mathbf{r}\,\left(\frac{\partial V\left[\delta g_{N}\right]}{\partial\mathbf{r}}\delta g_{N}\right).
\]
}After time averaging, we could have obtained the path large deviations
by studying the large deviations of the time averaged current. Alternatively,
we can rephrase our large deviation principle as a large deviation
principle for the current, through a change of variable. This is the
subject of this appendix. 

The conservative nature of the dynamics is visible because the large
deviation Hamiltonian $H$ (\ref{eq:final_H}) does depend on the
conjugate momentum $p$ only through its gradient $\partial p/\partial\mathbf{v}$.
We define $\tilde{H}$ as $\tilde{H}\left[f,\partial p/\partial\mathbf{v}\right]=H\left[f,p\right].$
We start from the definition of the large deviation Lagrangian 
\[
L\left[f,\dot{f}\right]=\text{Sup}_{p}\left\{ \int\text{d}\mathbf{v}\,\dot{f}p-H[f,p]\right\} .
\]
Writing $\dot{f}$ as the divergence of a current $\dot{f}+\partial/\partial\mathbf{v}\cdot\mathbf{j}=0$,
we have 
\[
L\left[f,\dot{f}\right]=\underset{\left\{ \mathbf{j}\mid\dot{f}+\frac{\partial}{\partial\mathbf{v}}\cdot\mathbf{j}=0\right\} }{\text{Sup}}\text{Sup}_{p}\left\{ -\int\text{d}\mathbf{v}\,p\frac{\partial}{\partial\mathbf{v}}\cdot\mathbf{j}-H[f,p]\right\} .
\]
Using $H\left[f,p\right]=\tilde{H}\left[f,\partial p/\partial\mathbf{v}\right]$,
and integrating by part, we have 
\[
L\left[f,\dot{f}\right]=\underset{\left\{ \mathbf{j}\mid\dot{f}+\frac{\partial}{\partial\mathbf{v}}\cdot\mathbf{j}=0\right\} }{\text{Sup}}\tilde{L}\left[f,\mathbf{j}\right]
\]
with
\[
\tilde{L}\left[f,\mathbf{j}\right]=\underset{\mathbf{E}}{\text{Sup}}\left\{ \int\text{d}\mathbf{v}\,\mathbf{j}\cdot\mathbf{E}-\tilde{H}[f,\mathbf{E}]\right\} .
\]
where $\mathbf{E}$ designates the conjugate quantity of the current
$\mathbf{j}$.

We thus have the large deviation principle
\begin{equation}
\mathbf{P}\left(\left\{ f_{N}(\tau)\right\} _{0\leq\tau\leq T}=\left\{ f(\tau)\right\} _{0\leq\tau\leq T}\right)\underset{N\rightarrow\infty}{\asymp}\text{e}^{-NL^{3}\underset{\left\{ \mathbf{j}\mid\dot{f}+\frac{\partial}{\partial\mathbf{v}}\cdot\mathbf{j}=0\right\} }{\text{Sup}}\int_{0}^{T}\text{d}\tau\,\tilde{L}\left[f,\mathbf{j}\right]}\text{e}^{-NI_{0}\left[f\left(\tau=0\right)\right]}.\label{eq:PGD_courant}
\end{equation}
We note that we can also write a large deviation principle for the
joint probability of the empirical density and the time averaged current
$\mathbf{j}_{N}\left(\tau\right)$

\textbf{
\[
\mathbf{P}\left(\left\{ f_{N}(\tau),\mathbf{j}_{N}\left(\tau\right)\right\} _{0\leq t\leq T}=\left\{ f(\tau),\mathbf{j}\left(\tau\right)\right\} _{0\leq t\leq T}\right)\underset{N\rightarrow\infty}{\asymp}\text{e}^{-N\mathcal{A}\left[f,\mathbf{j}\right]}\text{e}^{-NI_{0}\left[f\left(\tau=0\right)\right]},
\]
}with
\[
\mathcal{A}\left[f,\mathbf{j}\right]=\begin{cases}
\begin{array}{ll}
L^{3}\int_{0}^{T}\text{d}\tau\,\tilde{L}\left[f,\mathbf{j}\right] & \text{if\,\,\,}\text{\ensuremath{\dot{f}}+\ensuremath{\frac{\partial}{\partial\mathbf{v}}\cdot}\ensuremath{\mathbf{j}}=0,}\\
+\infty & \text{otherwise}.
\end{array}\end{cases}
\]
\textbf{ }
\begin{acknowledgements}
We warmly thank Oleg Zaboronski for first teaching us the Szegö--Widom
theorem, used in this paper. We thank Julien Barré, Charles-Edouard
Bréhier, Jules Guioth, and Julien Reygner for useful comments on our manuscript. The
research leading to this work was supported by a sub-agreement from
the Johns Hopkins University with funds provided by Grant No. 663054
from Simons Foundation. Its contents are solely the responsibility
of the authors and do not necessarily represent the official views
of Simons Foundation or the Johns Hopkins University. We thank the two anonymous reviewers for their useful recommendations,
which helped us to improve our first version of our manuscript.
\end{acknowledgements}



\bibliographystyle{spmpsci}
\bibliography{ref-lenard-balescu}

%\begin{thebibliography}{}
%%
%% and use \bibitem to create references. Consult the Instructions
%% for authors for reference list style.
%\end{thebibliography}

\end{document}